\documentclass[a4paper,fleqn]{cas-dc}

\usepackage[numbers,sort&compress]{natbib}
\usepackage{subfigure}
\usepackage{tikz}
\usepackage{adjustbox}
\usepackage{lipsum}
\usepackage{algorithm2e}
\usepackage{threeparttable, tablefootnote}
\usepackage{subfigure}
\usepackage{wrapfig}
\usepackage{adjustbox}
\usepackage{float}
\usepackage{textcomp}
\usepackage{soul}
\usepackage{ulem}
\usepackage{array}
\usepackage{amsmath}

\def\tsc#1{\csdef{#1}{\textsc{\lowercase{#1}}\xspace}}
\tsc{WGM}
\tsc{QE}
\tsc{EP}
\tsc{PMS}
\tsc{BEC}
\tsc{DE}

\begin{document}
\let\WriteBookmarks\relax
\def\floatpagepagefraction{1}
\def\textpagefraction{.001}
\shorttitle{}
\shortauthors{Islas et~al.}

\title [mode = title]{Biomass dust explosions: CFD simulations and venting experiments in a 1 m\textsuperscript{3} silo}




\author[1]{{Alain Islas}}[]


\address[1]{Department of Energy, University of Oviedo - 33203 Gijón, Asturias, Spain}

\author[1]{{Andrés Rodríguez Fernández}}[]

\author[2]{{Covadonga Betegón}}[%
   ]


\address[2]{Department of Construction and Manufacturing Engineering, University of Oviedo - 33203 Gijón, Asturias, Spain}

\author[3]{{Emilio Martínez-Pañeda}}[]

\address[3]{Department of Civil and Environmental Engineering, Imperial College London - London, SW7 2AZ, United Kingdom}

\author[1]{{Adrián Pandal}}[orcid=0000-0001-6006-2199]
\cormark[1]
\ead{pandaladrian@uniovi.es}

\cortext[cor1]{Corresponding author:}

\begin{abstract}
This study presents CFD simulations of biomass dust explosions in a newly developed experimental 1 m\textsuperscript{3} silo apparatus with variable venting, designed and fabricated to operate similarly to the explosivity test standards. The aim of the study is to validate a CFD model under development and investigate its capability to capture the transient effects of a vented explosion. The model is based on OpenFOAM and solves the multiphase (gas-particle) flow using an Eulerian-Lagrangian approach in a two-way regime. It considers the detailed thermochemical conversion of biomass, including moisture evaporation, devolatilization, and char oxidation, along with the homogeneous combustion of gases, turbulence, and radiative heat transfer. The explosion is analyzed in all stages, i.e., dust cloud dispersion, ignition, closed explosion, and vented explosion. The results indicate excellent agreement between the CFD model and experimental tests throughout the sequence. Our findings highlight the critical role of particle size in dust cloud distribution and pre-ignition turbulence, which significantly influences flame dynamics and the explosion itself. This model shows great promise and encourages its application for future investigations of biomass dust explosions in larger-scale geometries, especially in venting situations that fall out of the scope of the NFPA 68 or EN 14491 standards, and to help design effective safety measures to prevent such incidents.
\end{abstract}


\maketitle

\section{Introduction}
\label{Section:Introduction}

As global efforts to achieve net zero emissions by 2050 continue, the demand for bio-energy has increased, making biomass combined heat \& power (CHP) an attractive option for greenhouse gas (GHG) abatement due to its CO\textsubscript{2} neutral characteristics \cite{yang2022role,tabriz2022biomass} and potential to become carbon negative if combined with carbon capture and storage (CCS) \cite{taipabu2022critical}. From biomass co-firing to dedicated routes, long-term fuel delivery concepts and contracts are essential for the successful operation of biomass power plants \cite{IEABioenergy}. However, the supply and availability of feedstock must be carefully considered to ensure continuous and stable power generation \cite{simbolotti2007biomass}. Unfortunately, experience has shown that dust explosions are a potential hazard that must be addressed through the implementation of necessary precautions to guarantee safe and reliable plant operation, particularly during fuel handling and storing phases \cite{Copelli2019329,ABUSWER201649}.\\


Dust explosions are a significant threat in power plants and other industrial facilities, posing a peril to worker safety and property damage \cite{santamaria2023characterization,YUAN201557}. These explosions can occur in a range of equipment, from silos and mills to conveyors and dust collection systems, and can be triggered by various sources, including hot surfaces, electrical sparks, and self-heating processes \cite{ABBASI20077}. While prevention and inherent safety measures are the primary means of reducing the hazards of dust explosions \cite{amyotte2009application}, it is often necessary to implement operational and dynamic risk assessments to better comprehend the probability of occurrence, the potential severity of dust explosions, and to look for mitigation solutions such as venting panels \cite{qiu2021experimental,wang2023duct}. However, determining the appropriate vent size remains a controversial issue \cite{eckhoff2005current,tascon2017design,huang2022vented}, despite the existence of established standards, e.g., the EN 14491 or the NFPA 68 codes \cite{NFPA68,standard2008dust}.\\


Likewise, as newly built plants scale up, more cost-efficient, high-volume storage solutions are needed to secure continuing plant operation. Although mammoth silos may seem an attractive option \cite{schott2004large}, they often fall outside the scope of these standards, highlighting the need for further research.\\

To address these challenges, besides the traditional dust explosion testing activities \cite{ZHOU2019144,LIN201948,liu2019explosion,SONG2020429,kuracina2021study,pietraccini2023study}, modeling research \cite{YANG202172,LI2018360,CLONEY2018215,CHAUDHARI2019192,PICO2020cfddpm,SKJOLD2005151} has emerged as an alternative to predict the consequences of dust explosions with reduced labor and capital. Especially, computational fluid dynamics (CFD) simulations can play a meaningful role in assessing risk analysis, providing a more nuanced understanding of explosion development and designing mitigation systems beyond the simplified scenarios considered by guidelines and standards. These tools have been successfully applied to the study of various aspects related to dust explosions, including dust cloud formation \cite{DIBENEDETTO2013cfd,waduge2017predicting,RANI201514,DISARLI2019JLP,SERRANO20211,ISLAS2022117033,portarapillo2020cfd_1m3vessel,portarapillo2021cfd_diam,portarapillo2022cfd_equipped,ren2022experimental,klippel2013investigations,klippel2014dust} and the determination of the explosion severity parameters \cite{SKJOLD2005151, LI2020cornstarch, PICO2020cfddpm, islas2022computational, PORTARAPILLO2021turb}. However, the practical application of CFD codes to large industrial settings requires a pragmatic approach that involves a compromise with accuracy and precision, and initial validation through repeated small-scale experiments is necessary \cite{KHAN2015116}.\\

In this paper, we aim to contribute to safety engineering and consequence analysis by presenting the next step in our efforts to develop a reliable computational tool for simulating dust explosions in industrial equipment. Specifically, we validate the performance of our previous CFD model \cite{ISLAS2022117033,islas2022computational} by revisiting it and conducting experiments on dust explosion venting. To do this, we designed a self-made silo with a capacity of 1 m\textsuperscript{3} that features adjustable venting, and we used it to perform biomass dust explosions. We based our test procedure on the EN 14034 \cite{EN14034} and ASTM E1226 \cite{ASTME1226} standards, and we constructed the silo based on the design references for standardized test vessels, including the 20L Siwek sphere and 1 m\textsuperscript{3} ISO chamber.\\

The purpose of this study is to gather experimental data and use it to validate our CFD model's ability to capture the transient behavior that occurs during the different stages of a dust explosion. These stages include: (1) dust cloud dispersion, (2) ignition, (3) pressure development and flame propagation, and (4) pressure relief. Our end goals for this research are two-fold: (1) to enhance the potential of CFD codes to accurately simulate biomass dust explosions and (2) to improve the accuracy of vent sizing calculations to reduce the risk of dust explosions in industrial settings.\\

\begin{figure*}[h]
    \centering
    \includegraphics[width=\textwidth]{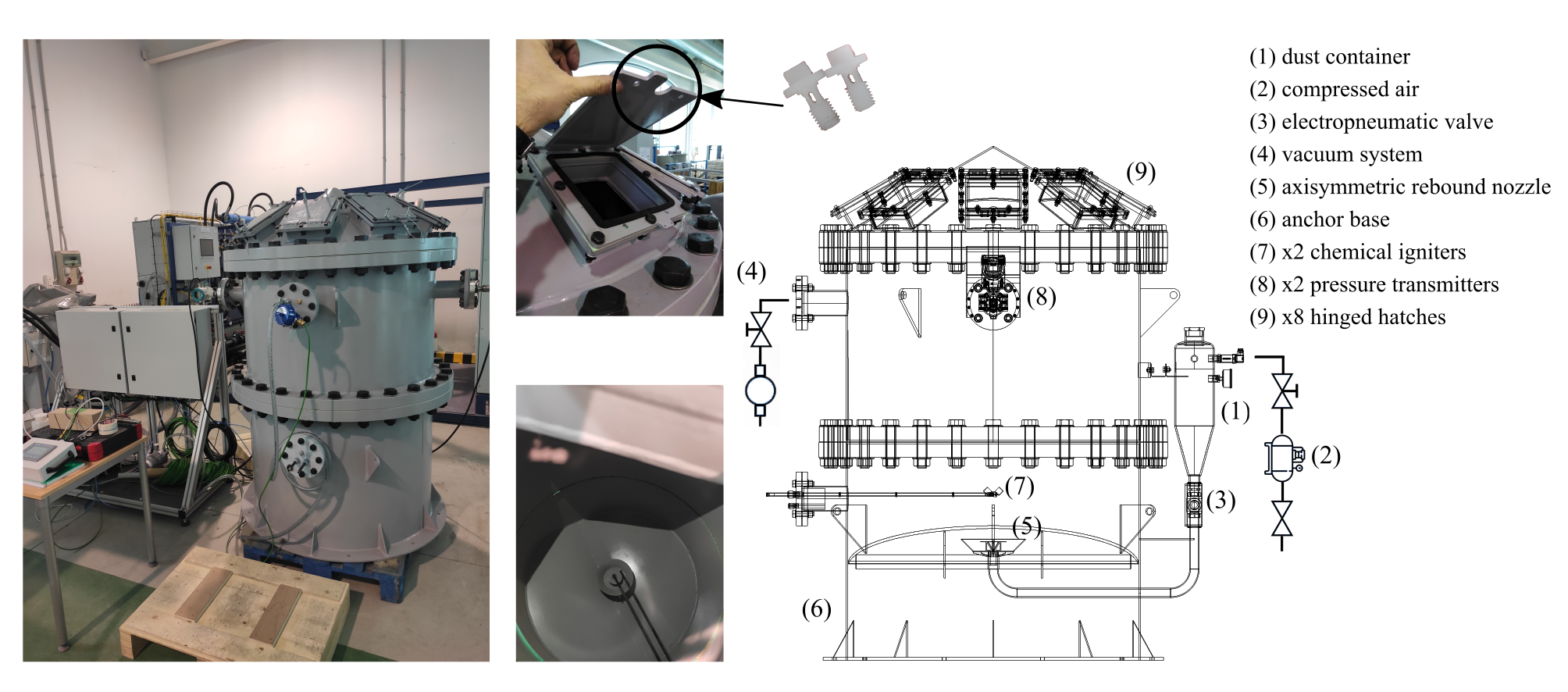}
    \caption{Layout of the 1 m\textsuperscript{3} silo apparatus}
    \label{Fig:1m3_Silo}
\end{figure*}

\section{Materials and methods}
\label{Section:Materials_and_methods}

\subsection{Experimental setup}
\label{Subsection:Experimental_setup}

A 1 m\textsuperscript{3}, pressure-resistant silo with adjustable venting was designed in partnership with PHB Weserhütte S.A. and the R\&D center IDONIAL in Asturias, Spain. The silo was manufactured in carbon steel ASME SA516-GR70, resistant up to 5 bar g overpressure and its dimensions were scaled down from a typical design of large-scale silo ( >10,000 m\textsuperscript{3}). The bottom is beveled to imitate a hopper design and the roof is cone-shaped. The vent openings consist of 8 hinged hatches (198x185 mm), distributed equiangularly on the roof, see Fig. \ref{Fig:1m3_Silo}. The venting area varies between 0.036 to 0.293 m\textsuperscript{2} representing up to a total venting efficiency of 30.64\%. The opening is regulated by polymer bolts specifically designed to withstand a 570 mbar g overpressure. The sizing of these bolts was calculated based on the tensile strength of the material, which required drilling the threaded shank to obtain the appropriate cross-section.\\

To create the dust cloud inside the test vessel, 
a system of pressurized air injection was employed for dispersing the dust sample into the silo. This system consists of a dust canister that 
has a volume of 5 L and a length-to-diameter ratio $L/D=3.6$. Its lower part is cone-shaped to facilitate dust outflow. As in the standards, the canister is pressurized up to 20 bar g, but the 1 m\textsuperscript{3} silo is vacuumed to -0.125 bar g prior the start of the dispersion process. This condition is important to ensure that the normal pressure at the start of the deflagration test is exactly 0.0 bar g. The air discharge is controlled by a Nordair\textsuperscript{\tiny\textregistered} U150 electropneumatic valve with an ATEX II 2GD actuator. The ignition delay time $t_{d}$ was set to 600 ms in all the experiments, matching the value used for the tests in the 1 m\textsuperscript{3} ISO chamber \cite{EN14034}.
To favor the radial spread of the dust, a new nozzle was designed. Specifically, an axisymmetric version of the traditional \textit{rebound nozzle} \cite{ASTME1226,MURILLO201854} was manufactured and installed at the bottom of the silo.\\

The dust cloud is ignited by means of 2x5 kJ Sobbe\textsuperscript{\tiny\textregistered} chemical igniters placed right above the dispersion nozzle and upheld by two slender rods. The igniters are fired oppositely at an angle of $45^\circ$ with respect to the horizontal. The pressure reading is recorded by two pressure transmitters Siemens\textsuperscript{\tiny\textregistered} Sitrans P320 positioned at the top and on opposite extremes of the cylindrical walls. The control and data acquisition system consists of a programmable logic controller (PLC) Siemens\textsuperscript{\tiny\textregistered} Simatic HMI and a videocamera. All the tests were conducted in the experimental test site of Applus+ TST\textsuperscript{\tiny\textregistered} (Tunnel Safety Testing, S.A.) in Asturias, Spain.

\subsection{Dust sample}
\label{Subsection:Dust_sample}

\begin{figure}[pos=h]
    \centering
    \includegraphics[width=0.45\textwidth]{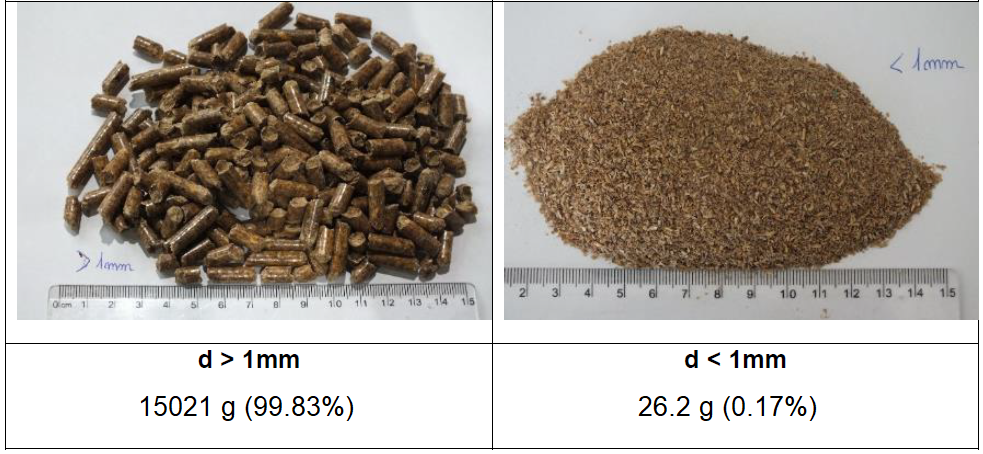}
    \caption{Biomass pellets and fine particulates in a commercial pellet bag}
    \label{Fig_3:pellets}
\end{figure}

The aim of this research is to study dust explosions with a representative sample found in industrial processes that manipulate pellets. The test sample is a commercial biomass from a local pellet manufacturer in Asturias, Spain, and is comprised of natural wood sub-products (saw dust, wood chips and debarked wood). The commercial pellets were received in a 15 kg bag format, whose percentage of fine particulates ($d<1\,\text{mm}$) was less than 1\%, see Fig. \ref{Fig_3:pellets}. As only a few grams could be used for the explosion tests, the pellets were ground in a gently-rotating ball mill to generate additional combustible dust. 

\begin{figure}[h]
    \centering
    \includegraphics[width=0.5\textwidth]{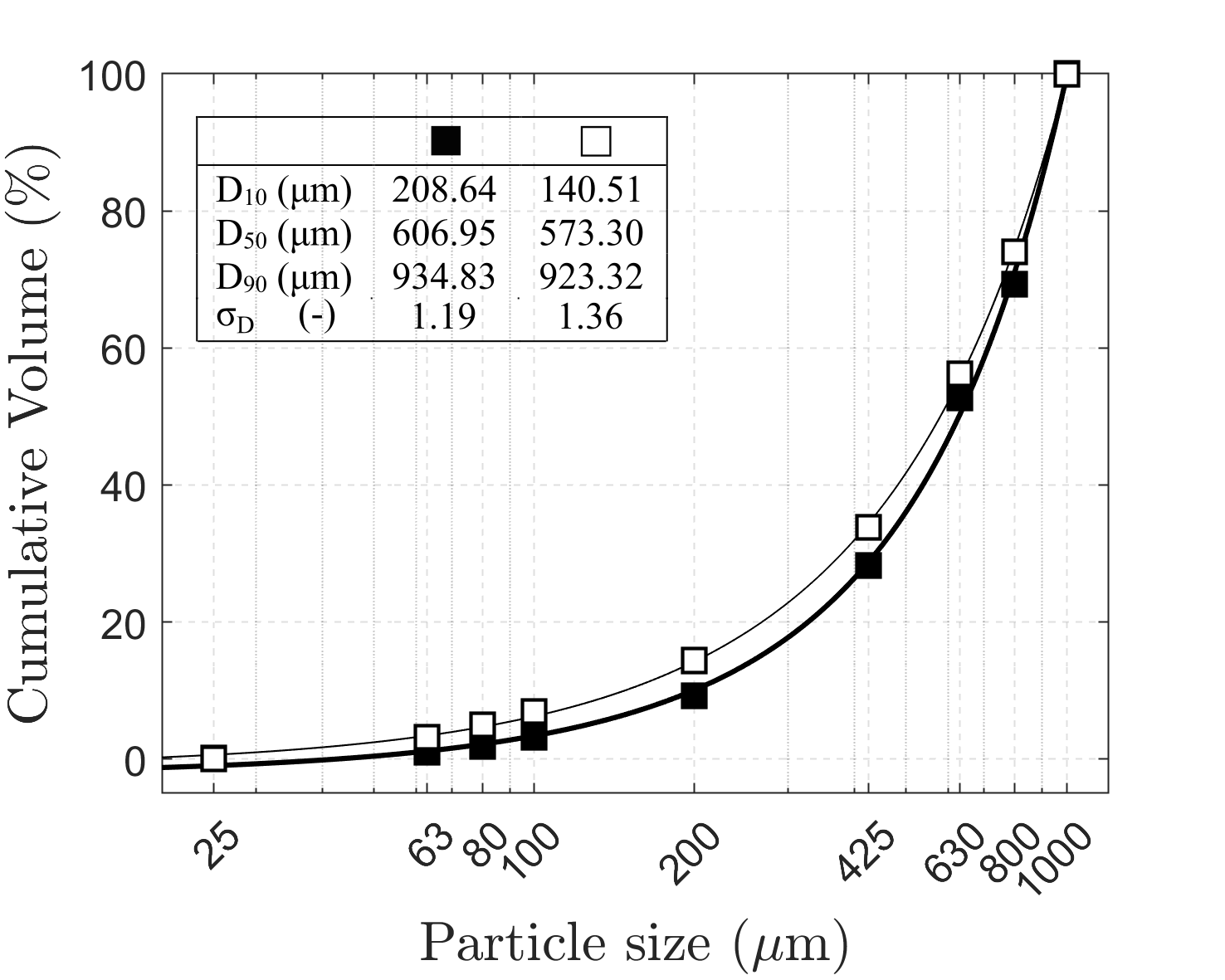}
    \caption{Particle size distribution of the fine particulates originally contained in a commercial pellet bag $\blacksquare$ and additional dust generated from pellet milling $\square$.}
    \label{Fig:PSD_comparison}
\end{figure}

The particle size distribution (PSD) of both the raw fine particulates in the bag and the post-milling samples was determined by Sieve analysis. The cumulative distributions and other size statistics are shown in Fig. \ref{Fig:PSD_comparison}.\\

As noted, the PSD generated by pellet milling contains slightly more fine particle diameters than the PSD of the raw dust. However, the difference is small as indicated by the polydispersity index. The polydispersity index $\sigma_{D}$ is a measure of the breadth in a size distribution \cite{merkus2009particle,tascon2018influence} and is calculated as

\begingroup
\small
\begin{flalign}
    \sigma_{D}=\frac{D_{90}-D_{10}}{D_{50}}&&
\end{flalign}
\endgroup

\noindent where the median $D_{50}$ and the $D_{10}$ and the $D_{90}$ values represent the 50\%, 10\% or 90\% point in the cumulative undersize PSD, respectively. A polydispersity index $\sigma_{d}\ll 1$ indicates a high homogeneity in particle size or a narrow PSD, while $\sigma_{D}\gg 1$ represents a heterogeneous particle size or a broad PSD \cite{CASTELLANOS2014331}. Considering that in industrial applications, the dust particles can vary in size largely, the polydispersity indices of both the original and ground PSDs are comparable. Moreover, the two PSD cover the same order of magnitude and the median varies in $\sim 5\%$. So, for practical purposes the former size distribution is considered to be representative of typical transporting, handling, and stacking activities of pellets.\\

The ultimate and proximate analysis, as well as the lower calorific value (LCV) of the sample were taken from the manufacturer's specifications sheet, see Table \ref{Table:Proximate_Ultimate_analysis}.

\begin{table}[H]
\caption{Ultimate and proximate analyses of the biomass sample}
\label{Table:Proximate_Ultimate_analysis}
\begin{tabular}{@{}lc@{}}
\toprule
Label                             & Pellets Asturias          \\ \midrule
Sample                            & Biomass pellets \\
                                  &  \\
\textit{Proximate analysis (wt. \% ar)} & \\
Fixed carbon                      & 14.16   \\
Volatile matter                   & 77.04   \\
Moisture                          & 8.33    \\
Ash                               & 0.47    \\
                                  &         \\
\textit{Ultimate analysis (wt. \% daf)}&    \\
C                                 & 50.25   \\
H                                 & 6.02    \\
O                                 & 43.45   \\
N                                 & 0.28    \\
Lower calorific value (MJ/kg)     & 18.83   \\ \bottomrule
\end{tabular}
\end{table}

\subsection{Ignition}
\label{Subsection:ignition}


When the ignition energy is too strong relative to the chamber size, it can cause a significant increase in pressure \cite{ZHEN1997317}. Several studies have shown that using a 10 kJ ignition source in a small volume such as the 20L Siwek sphere, leads to an overdriving effect that cannot be ignored \cite{GOING2000209,cashdollar1993,TAVEAU2017348,KUAI2011302}. In closed vessel testing, the overdriving effect can lead to an overestimation of the explosion severity parameters \cite{CLONEY20131574,szabova2021influence}, particularly the deflagration index $K_{st}$ and the minimum explosive concentration (MEC).\\ 





\begin{figure}
    \centering
    \includegraphics[width=0.5\textwidth]{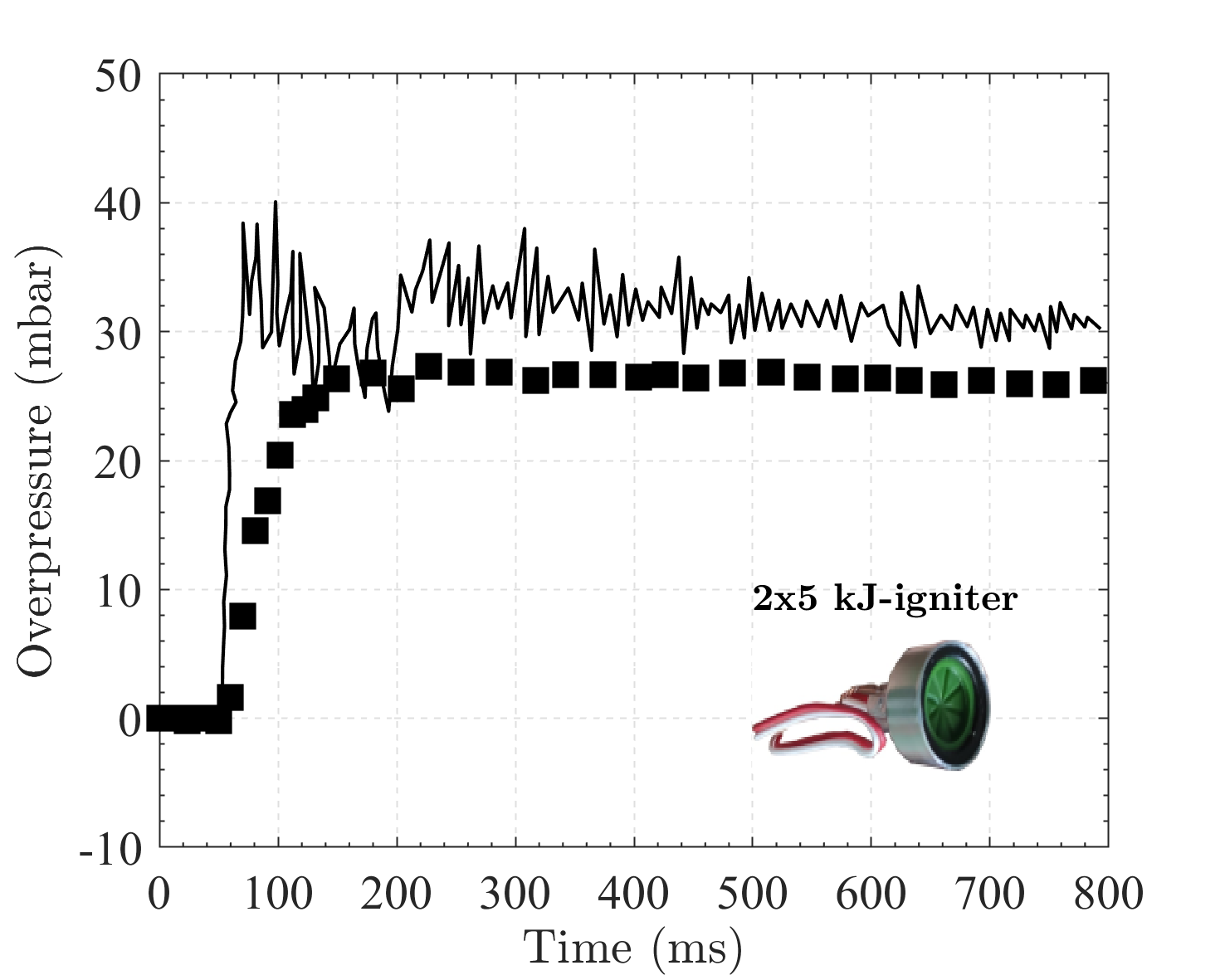}
    \caption{Comparison of the overpressure produced by 2x5 kJ igniters inside the 1 m\textsuperscript{3} silo apparatus ($\blacksquare$) and the 1 m\textsuperscript{3} ISO chamber (\textminus) (Zhen \& Leuckel, 1997 \cite{ZHEN1997317}).}
    \label{Fig:Overpressure_igniters}
\end{figure}

 Each of the 5kJ Sobbe\textsuperscript{\tiny\textregistered} chemical igniters is charged with 1.2 g of a pyrotechnic powder mixture of 40\% w.t. zirconium, 30\% barium nitrate, and 30\% barium peroxide \cite{HERTZBERG1988303}. They are activated electrically with internal fuse wires. To determine whether an overdriving effect takes place in the 1 m\textsuperscript{3} silo, we conducted a blank test experiment. The pressure-time trace developed by the 2x5 kJ igniters alone was measured in the closed silo and free of any combustible dust. Fig. \ref{Fig:Overpressure_igniters} presents a comparison between the resulting overpressures in the 1 m\textsuperscript{3} silo and a 1 m\textsuperscript{3} ISO chamber from the literature \cite{ZHEN1997317}.\\


 Clearly, there is jump-like behavior during the first moments after the igniters are triggered. This is because the igniters deliver their energy in very short times ($\sim 10\,\text{ms}$) \cite{cashdollar1993}. The maximum overpressure is registered as approximately 30 mbar g and matches reasonably well the same time-dependency as the experiment in the 1 m\textsuperscript{3} ISO chamber. Moreover, the 30 mbar g overpressure is consistent with the values reported in other studies \cite{TAVEAU2017348,cashdollar1993}. According to data collected from blank test experiments in the 20L Siwek sphere, the pressure increase due to 10kJ igniters can vary between 0.8 and 1.6 bar \cite{FUMAGALLI201893,ZHAO2020116401,portarapillo2021ignit}. Therefore, when compared to the overpressure in any 1 m\textsuperscript{3} volume, the overdriving effect can be safely regarded as negligible.



\section{Gas and particle phase modeling}
\label{Section:Gas_and_particle_phase_modeling}
In this work, the vented biomass dust explosions were simulated in the 1 m\textsuperscript{3} silo by employing our customized version of OpenFOAM's coalChemistryFoam code \cite{islas2022computational}.
This CFD code is a transient solver of two-phase (gas-solid) flow suitable to model compressible flow with turbulence, combustion, chemical reactions and radiative heat transfer. The solver uses an Eulerian-Lagrangian method to solve the particle-laden flow within a two-way coupling regime. Source terms are computed to represent the exchange of
mass, momentum, energy and chemical species between the two phases. The Lagrangian framework allows for a detailed analysis of biomass burning, including modeling of sensible heating and thermochemical conversion of biomass. To reduce the computational burden, physical particles are replaced with computational parcels, which group together particles with similar properties and whose extensive properties are scaled by a number density.

\subsection{Gas phase governing equations}

In CFD simulations of dust explosions, the Reynolds-Averaged Navier Stokes (RANS) closure is often used. The gas phase governing equations consist of the Reynolds-averaged mass, momentum, energy and species transport equations. The mass transport is

\begingroup
\small
\begin{flalign}
    \frac{\partial \bar{\rho}}{\partial t} + \frac{\partial}{\partial x_{i}}\left(\bar{\rho} \tilde{u}_{i}\right)= \Gamma_{i}&&
    \label{Eqn:mass_transport}
\end{flalign}
\endgroup

\noindent where the overbar denotes that the scalar is Reynolds-averaged and the tilde denotes density-weighted time averaged or \textit{Favre-averaged}. As reacting particles can exchange mass with the gas phase, the source term $\Gamma_{i}$ is included in Eq. (\ref{Eqn:mass_transport}) to account for the fluid/particle interaction. The momentum transport equations are

\begingroup
\small
\begin{flalign}
    \frac{\partial}{\partial t}\left(\bar{\rho} \tilde{u}_{i}\right) + \frac{\partial}{\partial x_j}\left(\bar{\rho} \tilde{u}_{i} \tilde{u}_{j}\right)= - \frac{\partial \bar{p}}{\partial x_j} + \frac{\partial \bar{\tau}^{ij}}{\partial x_j}+\frac{\partial}{\partial x_j}\left(-\bar{\rho} \widetilde{u_i^{\prime} u_j^{\prime}}\right)
    +\bar{\rho} g_i + \Lambda_{i}&&
    \label{Eqn:momentum_transport}
\end{flalign}
\endgroup

\noindent where the Reynolds stress term is calculated using the Bousinessq hypothesis $-\bar{\rho} \widetilde{u_i^{\prime} u_j^{\prime}}=2\mu_{t}\widetilde{S_{ij}}-\frac{2}{3}\bar{\rho}k$. The standard $k-\varepsilon$ turbulence model is used to determine the eddy viscosity $\mu_{t}=\bar{\rho}C_{\mu}k^2/\varepsilon$, where $k$ is the turbulent kinetic energy and $\varepsilon$ is the turbulence dissipation rate. $k$ and $\varepsilon$ are modeled using the following transport equations

\begingroup
\small
\begin{flalign}
    \frac{\partial }{\partial t}\left(\bar{\rho}k\right) + \frac{\partial}{\partial x_{i}}\left(\rho \tilde{u_{i}}k\right)=\frac{\partial}{\partial x_{i}}\left[\left(\mu+\frac{\mu_{t}}{\sigma_{k}}\right)\frac{\partial k}{\partial x_{i}}\right]+P_{k}-\bar{\rho}\varepsilon&&
    \label{Eqn:turbulentKineticEnergy_transport}
\end{flalign}
\endgroup

\begingroup
\small
\begin{flalign}
    \frac{\partial}{\partial t}\left(\bar{\rho}\varepsilon\right) + \frac{\partial}{\partial x_{i}}\left(\rho \tilde{u_{i}}\varepsilon\right)=\frac{\partial}{\partial x_{i}}\left[\left(\mu+\frac{\mu_{t}}{\sigma_{\varepsilon}}\right)\frac{\partial \varepsilon}{\partial x_{i}}\right]+C_{\varepsilon1}\frac{\varepsilon}{k}P_{k}-C_{\varepsilon2}\bar{\rho}\frac{\varepsilon^2}{k}&&
    \label{Eqn:turbulenceDissipationRate_transport}
\end{flalign}
\endgroup

Again, a source term $\Lambda_{i}$ is included in Eq. (\ref{Eqn:momentum_transport}) to represent the momentum exchange due to particles. The enthalpy transport equation is

\begingroup
\small
\begin{flalign}
\frac{\partial}{\partial t}\left(\bar{\rho} \widetilde{h}\right) + \frac{\partial}{\partial x_i}\left(\bar{\rho} \tilde{u}_{i} \widetilde{h}\right) = \frac{D \bar{p}}{D t} - \frac{\partial \bar{q_i}}{\partial x_i} + \overline{\tau^{ij}\frac{\partial u_i}{\partial x_j}}+\Theta_{i}&&
\label{Eqn:enthalpy_transport}
\end{flalign}
\endgroup

\noindent where $\Theta_{i}$ is a source term that accounts for the combined effect of: (1) the homogeneous gas phase reactions, (2) the enthalpy exchange due to the thermochemical conversion of the biomass particles, and (3) the radiative heat transfer. The species transport equation is

\begingroup
\small
\begin{flalign}
    \frac{\partial}{\partial t} \left(\bar{\rho} \widetilde{Y}_{k}\right) + \frac{\partial}{\partial x_{i}} \left(\bar{\rho} \tilde{u_{i}}\widetilde{Y}_{k}\right) = \frac{\partial}{\partial x_{i}}\left(\bar{\rho} \overline{D}_{k}\frac{\partial \widetilde{Y}_{k}}{\partial x_{i}}\right) + \overline{\dot{\omega}}_{k} + \Phi_{k}&&
\end{flalign}    
\endgroup

\noindent where $\widetilde{Y}_{k}$ is the mass fraction of species $k$ in the gas mixture and $\overline{\dot{\omega}}_{k}$ is the chemical reaction rate. The source term $\Phi_{k}$ represents the species released/consumed by the particle devolatilization and char conversion.\\

\subsection{Particle governing equations}

In biomass dust explosions, the interaction between the particles and the surrounding medium is through mass and momentum exchange and heat transfer. In the CFD model, each biomass particle is a reactive multi-phase entity, whose content of liquid, gaseous and solid matter is based on the proximate analysis. The mass conservation for each particle is written as

\begingroup
\small
\begin{flalign}
    \frac{dm_{p}}{dt}=\dot{m}_{moisture}+\dot{m}_{volatiles}+\dot{m}_{char}&&
\end{flalign}
\endgroup

\noindent where $\dot{m}_{moisture}$, $\dot{m}_{volatiles}$, $\dot{m}_{char}$ denote the rate of evaporation, devolatilization, and char oxidation. After all the reactive content is depleted, the biomass is reduced to an inert ash particle.  Along its entire thermal history, the particle temperature is obtained from the energy conservation

\begingroup
\small
\begin{flalign}
    m_{p}C_{p}\frac{dT_{p}}{dt} = \pi d_{p}k_{g}\text{Nu}\left(T_{\infty}-T_{p}\right) + \frac{dm_{p}}{dt}\Delta H + \pi d_{p}^2\varepsilon_{0}\sigma\left(\theta_{R}^4-T_{p}^4\right)&&
\label{Eqn:Particle_energy_equation}
\end{flalign}
\endgroup

\noindent where $\frac{dm_{p}}{dt}$ and $\Delta H$ denote the rate of mass consumption within a particle and its associated latent heat due to one of the three mechanisms 

\begingroup
\small
\begin{flalign}
\frac{dm_{p}}{dt} = 
\left\{
    \begin{array}{lr}
        \pi d_{p}D_{0}\text{Sh}\left(\frac{p_{sat,T}}{RT_{m}}-X_{w}\frac{p}{RT_{m}}\right)M_{w} & \text{evaporation}\\
        -k\left(T\right)\left(m_{p}-\left(1-f_{\text{VM}_{0}}\right)m_{p_{0}}\right) & \text{devolatilization}\\
        -\pi d_{p}^2 p_{o}\left(\frac{1}{R_{diff}}+\frac{1}{R_{kin}}\right)^{-1} & \text{char oxidation}
    \end{array}
    \right.&&
\label{Eqn:Particle_Evaporation_Devolatilization_Charoxidation}
\end{flalign}
\endgroup



The heat and mass transfer numbers, $\text{Nu}$ and $\text{Sh}$ are found using the Ranz-Marshall correlations for spherical particles \cite{ranz1952evaporation}. Eq. (\ref{Eqn:Particle_Evaporation_Devolatilization_Charoxidation}) states that biomass combustion can be seen as a three-stage, sequential process: (1) evaporation of moisture, (2) thermal cracking of biomass into light gases, and (3) the heteregenous conversion of char. The evaporation of moisture consists of the endothermic phase change of liquid water contained within the particle into water vapor that is added to the gas phase. The devolatilization or thermal cracking of biomass is the release of volatile gases that are further combusted in the gaseous phase. Contrarily to other solid fuels (e.g., coal), the overall heat release of biomass samples is dominated by the combustion of these gases. For example, the volatile matter in Pellets Asturias represents more than 75\% of the total mass, see Table. \ref{Table:Proximate_Ultimate_analysis}. The remaining char is burned by the heteregeneous reaction with oxygen.\\

Depending on various characteristics, e.g., the heating rate, particle residence time or particle temperature, the gas species composition during devolatilization can be quite diverse. For the sake of model simplification, the volatiles are represented as a postulate substance $\text{C}_{x}\text{H}_{y}\text{O}_{z}$, whose x, y or z subscripts are calculated from the ultimate and proximate analysis. During devolatilization each biomass particle breaks down into the following 4 light gases \cite{sami2001co,di2008modeling,neves2011characterization}

\begingroup
\small
\begin{flalign}
    \text{C}\textsubscript{x}\text{H}\textsubscript{y}\text{O}\textsubscript{z}& \xrightarrow[]{k_{v}} \nu_{1}^{\prime\prime} \text{CO} + \nu_{2}^{\prime\prime} \text{CO\textsubscript{2}} + \nu_{3}^{\prime\prime} \text{CH\textsubscript{4}} + \nu_{4}^{\prime\prime} \text{H\textsubscript{2}} \label{Eqn:Mass_conservation_biomass} && \\
    \text{LCV}\textsubscript{VM} &= \sum_{i=1}^{4}Y_{i}\times\Delta H_{R,i}
\end{flalign}
\endgroup

 \noindent where the lower calorific value (LCV) of volatiles VM is found assuming that the LCV of biomass can be split into the combustion of its separate elements \cite{ansys2011ansys} 

\begingroup
\small
\begin{flalign}
    \text{LCV}_\text{biomass}&= Y_\text{VM}^{\text{daf}}\times\text{LCV}_\text{VM} + Y_\text{FC}^{\text{daf}}\times\text{LCV}_\text{FC}&&
    \label{Eqn:Energy_balance_biomass}
\end{flalign}
\endgroup

Under these considerations, the postulate volatile substance is $\text{C}_{1.03}\text{H}_{2.13}\text{O}_{0.97}$ and the stoichiometric coefficients $\nu_{i}^{\prime\prime}$ in Eq. (\ref{Eqn:Mass_conservation_biomass}) are 0.07, 0.44, 0.51, and 0.03 for CO, CO\textsubscript{2}, CH\textsubscript{4}, H\textsubscript{2} respectively. In all simulations, these gases are combusted following the 4-step reaction mechanism proposed by Jones \& Lindstedt \cite{joneslindstedt1988,islas2022computational}.\\
 
The kinematics of the particles is governed by Newton's 2nd law

\begingroup
\small
\begin{flalign}
    \frac{du_{p_{i}}}{dt}=\frac{18\mu}{\rho d_{p}^{2}}\frac{C_{D}\text{Re}_{p}}{24}\left(u_i-u_{p_{i}}\right)+g_i\left(1-\frac{\rho}{\rho_p}\right)&&
    \label{Eqn:particle_force_balance}
\end{flalign}
\endgroup

\begingroup
\small
\begin{flalign}
C_{D} = 
\left\{
    \begin{array}{lr}
        0.424 & \text{Re}_{p}> 1000\\
        \frac{24}{\text{Re}_{p}}\left(1+\frac{1}{6}\text{Re}_{p}^{2/3}\right) & \text{Re}_{p}\leq 1000
        \label{Eqn:Drag_coefficient}
    \end{array}
    \right. &&
\end{flalign}
\endgroup

\noindent where the RHS terms of Eq. (\ref{Eqn:particle_force_balance}) represent all the forces acting on the particle, namely drag, gravity and buoyancy. The drag factor is determined by the correlation for spherical particles proposed by Putnam \cite{putnam1961integratable}, Eq. (\ref{Eqn:Drag_coefficient}). Moreover, we use a stochastic dispersion approach to include the effect of instantaneous turbulent velocity fluctuations on the particle trajectories. With a given $u_{p_{i}}$ the position of the particle is computed by integrating the equation $\frac{dx_{p_{i}}}{dt}=u_{p_{i}}$. \\

Our customized solver includes more comprehensive submodels for the simulation of the biomass devolatilization and radiative heat transfer phenomena. Specifically, it uses the BioCPD model \cite{fletcher2012prediction,CPDfletcher2020} to determine the devolatilization kinetics of biomass samples at elevated heating rates and uses Mie theory calculations to estimate the radiative properties of particles. Moreover it uses a
a dry/wet \textit{weighted-sum of gray gase model} (WSGGM) model to calculate the absorption coefficient of the gaseous mixture \cite{kangwanpongpan2012new,smith1982evaluation}. For a detailed description of the complete method and the other submodels, the reader is referred to our previous works \cite{ISLAS2022117033,islas2022computational}.

\subsection{Computational grids}
\label{Subsection:computational_grids}

\begin{figure*}
    \centering
    \includegraphics[width=0.8\textwidth]{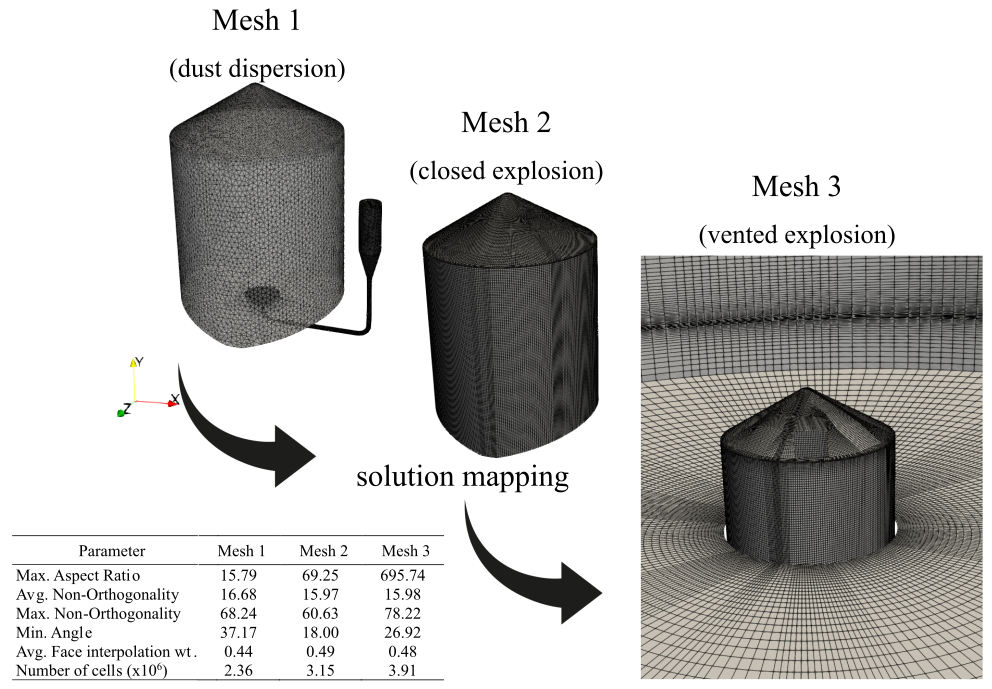}
    \caption{Layout of the computational grids and associated quality metrics.}
    \label{Fig:Mesh}
\end{figure*}

In order to simulate the full range of stages in a dust explosion (including dispersion, explosion, and venting), three separate computational grids were employed, see Fig. \ref{Fig:Mesh}. Mesh 1 corresponded to the entire silo, encompassing both the dispersion system and the silo itself. Mesh 2, on the other hand, focused solely on the inner region of the silo, excluding the dispersion system. Finally, mesh 3 was created as an exact copy of mesh 2, but with additional cells to represent the far field region. All grids were manually constructed using the ANSYS ICEM meshing software and subsequently converted to OpenFOAM format for simulation purposes.

\begin{table}[h]
\caption{Boundary conditions in the CFD model}
\label{Table:Boundary_conditions}
\begin{tabular}{@{}lll@{}}
\toprule
\multicolumn{1}{c}{Field} & \multicolumn{1}{c}{Walls} & \multicolumn{1}{c}{Far field} \\ \midrule
$p\,\left[\text{Pa}\right]$                & \texttt{zeroGradient}              & \texttt{101 325}                       \\
$\tilde{u}_{i}\,\left[\text{m/s}\right]$               & \texttt{noSlip}                    & \texttt{pressureInletOutletVelocity}   \\
$T\,\left[\text{K}\right]$                 & \texttt{zeroGradient}              & \texttt{zeroGradient}                  \\
$\tilde{Y}_{k}\,\left[\text{-}\right]$                  & \texttt{zeroGradient}              & \texttt{zeroGradient}                  \\
$k\,\left[\text{m}\textsuperscript{2}\,\text{s}\textsuperscript{-2}\right]$             & \texttt{kqRWallFunction}           & \texttt{zeroGradient}                  \\
$\varepsilon\,\left[\text{m}\textsuperscript{2}\,\text{s}\textsuperscript{-3}\right]$       & \texttt{epsilonWallFunction}       & \texttt{zeroGradient}                  \\ \bottomrule
\end{tabular}
\end{table}

\begin{table}[]
\caption{Initial conditions}
\label{Table:Initial_conditions}
\begin{tabular}{@{}llll@{}}
\toprule
\multirow{2}{*}{Field} & \multicolumn{1}{c}{Mesh 1}                                                                     & \multicolumn{1}{c}{Mesh 2 - Mesh 3} & \multicolumn{1}{c}{Mesh 3}                                                                     \\
                       & \multicolumn{1}{c}{(1 m\textsuperscript{3} silo)}                                                                 & \multicolumn{1}{c}{(1 m\textsuperscript{3} silo)}      & \multicolumn{1}{c}{(Far field)}                                                                \\ \midrule
$p\,\left[\text{Pa}\right]$                      & \begin{tabular}[c]{@{}l@{}}$p_{s}=87.5\times 10^{3}$ \\ $p_{c}=21\times 10^{6}$\end{tabular}                    & mapped                              & 101 325                                                                                         \\
 & & & \\
$\tilde{u}_{i}\,\left[\text{m}\,\text{s}\textsuperscript{-1}\right]$                      & 0                                                                                              & mapped                              & 0                                                                                              \\
$T\,\left[\text{K}\right]$                      & 293                                                                                            & mapped                              & 293                                                                                            \\
 & & & \\
\begin{tabular}[c]{@{}l@{}}$\tilde{Y}_{\text{O}\textsubscript{2}}\,\left[\text{-}\right]$\\ $\tilde{Y}_{\text{N}\textsubscript{2}}\,\left[\text{-}\right]$\\ $\tilde{Y}_{\text{other}}\,\left[\text{-}\right]$\end{tabular}                      & \begin{tabular}[c]{@{}l@{}} 0.2329 \\ 0.7671 \\ 0.0 \end{tabular}   & mapped                              & \begin{tabular}[c]{@{}l@{}} 0.2329 \\ 0.7671 \\ 0.0 \end{tabular} \\
 & & & \\
$k\,\left[\text{m}\textsuperscript{2}\,\text{s}\textsuperscript{-2}\right]$                      & 1                                                                                              & mapped                              & 1                                                                                              \\
$\varepsilon\,\left[\text{m}\textsuperscript{2}\,\text{s}\textsuperscript{-3}\right]$                      & 117                                                                                            & mapped                              & 117                                                                                            \\ \bottomrule
\end{tabular}
\end{table}

\subsection{Solution strategy}
\label{Subsection:Solution strategy}
The vented dust explosion was simulated in 3 stages, namely: (1) dust dispersion, (2) explosion in closed silo, and (3) vented explosion.

\begin{enumerate}
    \item \textbf{Dust dispersion:} the dust particles are initially placed in the dust container at stagnant conditions and the pressure field is initialized accordingly (i.e. 0.875 bar a in the silo and 21 bar a in the dust canister). The ensuing pressure gradient drives the particles from the canister to the silo. The dust injection is simulated for an ignition delay time $t_{d}=600 \,\text{ms}$. Right afterwards, the case is stopped and all the Eulerian and Lagrangian fields are mapped from mesh 1 to mesh 2.
    
    \item \textbf{Explosion in closed silo:} starting from the cold flow solution, the reactive features of the solver (combustion, chemistry, radiation, etc\ldots) are switched on. The ignition mechanism is activated and the dust cloud starts burning. In these simulations, the chemical igniters are again represented by a 10kJ enthalpy source term distributed over a kernel sphere of $r=13\,\text{cm}$ \cite{islas2022computational} placed right above the axisymmetric rebound nozzle. During the run-time, the instantaneous pressure is monitored each time-step as the weighted-area-average value on the roof surface. The simulation is stopped as soon as the monitor hits the rupture pressure of the polymer bolts, i.e. $p_{stat}=1.570\,\text{mbar a}$. Next, the Eulerian and Lagrangian fields are mapped from mesh 2 to mesh 3.
    
    \item \textbf{Vented explosion:} once mesh 3 is initialized with the latest reactive solution, the boundary condition at the venting areas are switched from \textit{walls} to \textit{interior cells}. The number of venting areas is changed based on the venting scenario. For the rest of simulation, the deflagration is allowed to escape to the surroundings and the pressure inside the 1 m\textsuperscript{3} silo decays. 
\end{enumerate}

\begin{figure*}
    \begin{tabular}{c|c}
    \subfigure[Time = -500 ms \label{Fig:Velocity_Streamlines_100ms}]{\includegraphics[width=0.50\textwidth]{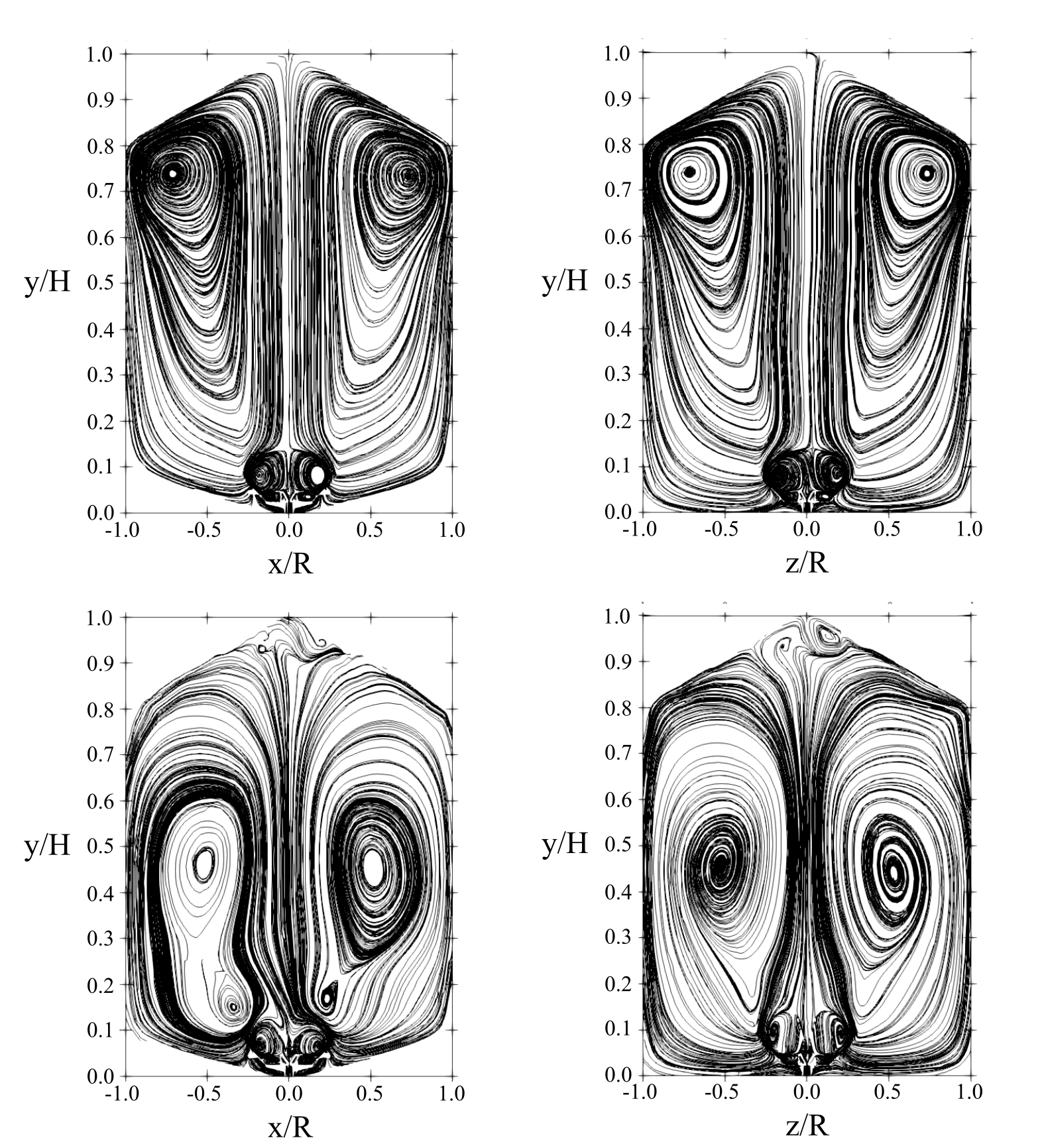}} & \subfigure[Time = 0 ms \label{Fig:Velocity_Streamlines_600ms}]{\includegraphics[width=0.50\textwidth]{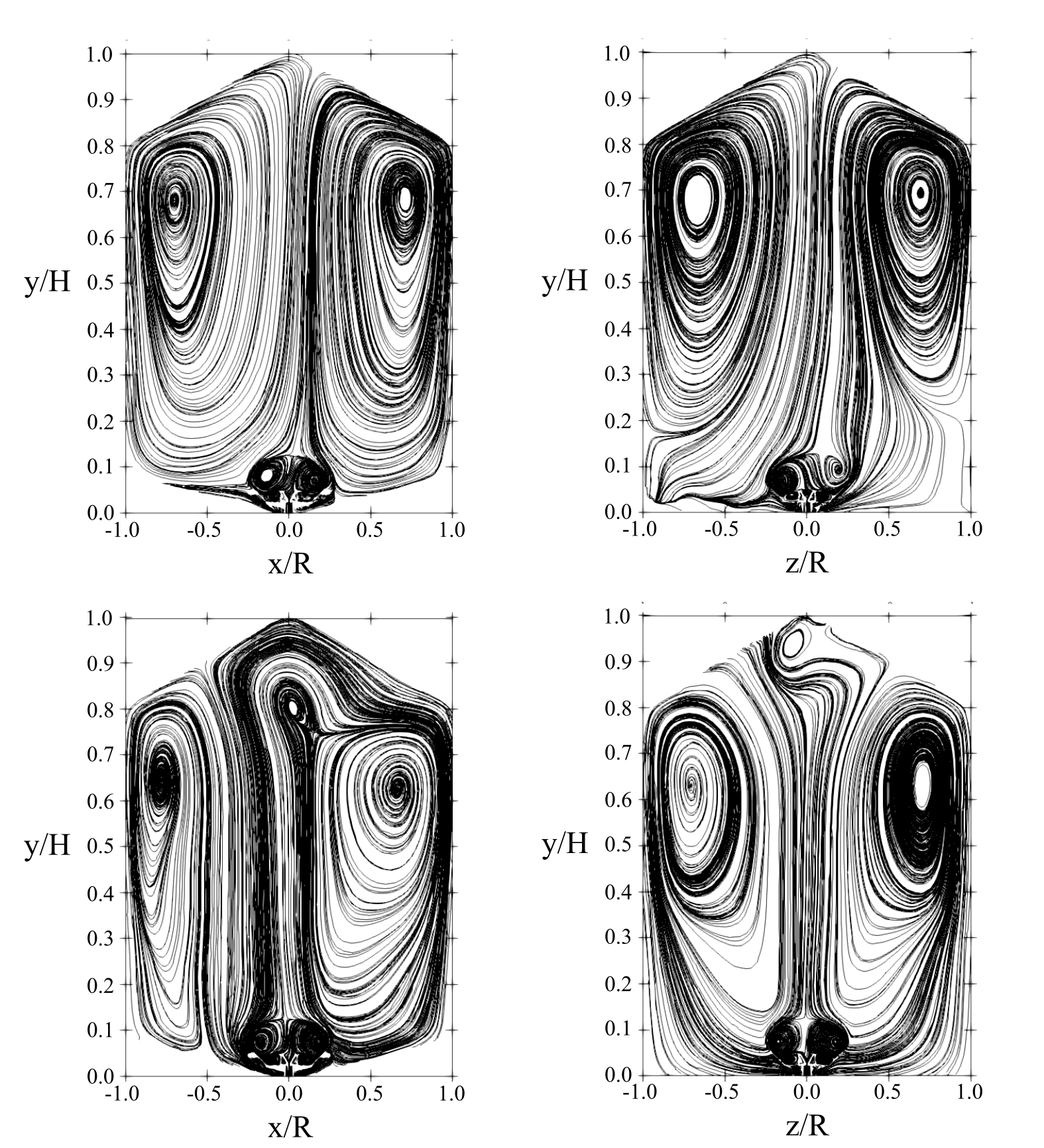}}
    \end{tabular}
    
    \caption{Velocity streamlines at selected times during the air blast. Dust-free flow (top row) and particle-laden flow (bottom row). Reference time (t=0 ms) at the ignition event.}
    \label{Fig:Velocity_Streamlines}
\end{figure*}

This approach enabled us to simulate the entire dust explosion using computational meshes tailored to specific purposes. Specifically, the grids 2 and 3 are structured and were meshed manually with topologies comprising hexagonal blocks. This helps ensure that all fluxes in the discretized equations, which involve numerous physics, pose high orthogonality and converge correctly. The boundary conditions and initialization settings of each stage of the simulation are provided in Table \ref{Table:Boundary_conditions} and Table \ref{Table:Initial_conditions}, respectively.\\

The conservation equations of the Eulerian phase were discretized using first-order upwind schemes and second-order difference schemes for the convective terms and diffusive terms, respectively. The gradients were evaluated using a cell-limited scheme scheme with cubic interpolation. The transient discretization was treated with a first-order Euler scheme with an adaptive time-stepping method to satisfy a Courant-Friedrich-Lewy (CFL) condition of CFL=1.0. The pressure-velocity coupling was solved by the PIMPLE algorithm with 3 correctors per time step. The flow residuals were set to $10^{-8}$ for continuity/pressure, and to $10^{-12}$ for momentum, energy, species and turbulence equations, respectively.

\section{Results and discussion}
\label{Section:Results_and_discussion}

\subsection{Dispersion system}
\label{Subsection:Dispersion_system}

\begin{figure}
    \centering
    \includegraphics[width=0.4\textwidth]{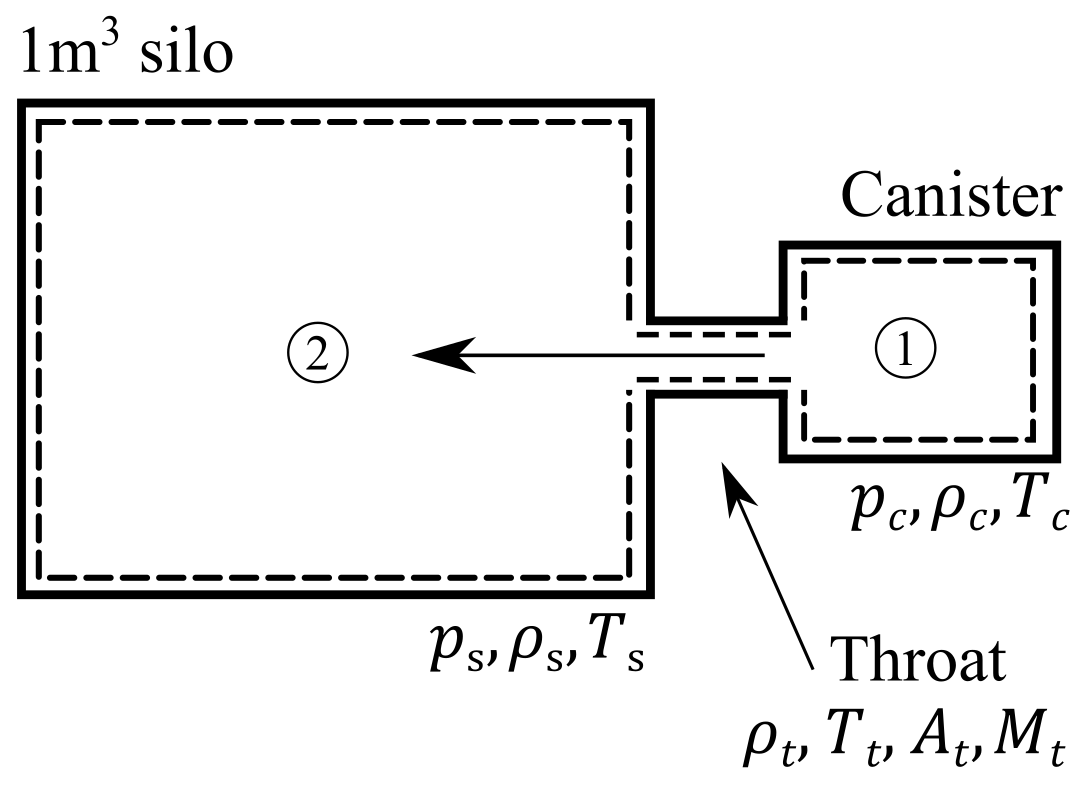}
    \caption{Charge and discharge of pressurized air between connected vessels}
    \label{Fig:charge_discharge}
\end{figure}

The transient behavior of the pressurized air injection can be estimated if we assume that the air is an ideal gas and that the discharge is modeled as a poly-tropic process ($p\text{\sout{\ensuremath{V}}}^{n}=C$), see Fig. \ref{Fig:charge_discharge}. If the pressure, temperature, and density in both the canister and 1 m\textsuperscript{3} silo are given at $t=0$, the pressure evolution $p_{C\text{\sout{\ensuremath{V}}},t+\Delta t}$ in either control volume $C\text{\sout{\ensuremath{V}}}$ 1 or 2 can be found as

\begingroup
\small
\begin{flalign}
    p_{C\text{\sout{\ensuremath{V}}},t+\Delta t}=p_{C\text{\sout{\ensuremath{V}}},t}\left(\frac{\rho_{C\text{\sout{\ensuremath{V}}},t+\Delta t}}{\rho_{C\text{\sout{\ensuremath{V}}},t}}\right)^{n}&& 
\end{flalign}
\endgroup

\noindent where $n$ is the poly-tropic exponent. The density at time $t+\Delta t$ is estimated by applying conservation of mass to the corresponding control volume

\begingroup
\small
\begin{flalign}
\frac{d \rho_{C\text{\sout{\ensuremath{V}}}}}{dt}=\pm \frac{\rho_{0}A_{t}M_{t}\left(\gamma R T_{0}\right)^{1/2}}{\text{\sout{\ensuremath{V}}}}\left(1+\frac{\gamma-1}{2}M_{t}^{2}\right)^{-\frac{\gamma+1}{2\left(\gamma-1\right)}}&&
\label{Eqn:mass_flow_rate_throat}
\end{flalign}
\endgroup

\noindent where a positive value represents a charging process and a negative value represents a discharging process. Since the vessels are connected, the mass leaving the canister equals the mass entering the silo. The RHS of Eq. (\ref{Eqn:mass_flow_rate_throat}) refers to the mass flow rate at the throat area $A_{t}$ considering the compressibility effects. The properties at the throat are calculated using isoentropic relations and assuming that the canister is at stagnant conditions ($T_{0}=T_{c}$ and $\rho_{0}=\rho_{c}$)

\begingroup
\small
\begin{flalign}
    \frac{T_{c}}{T_{t}}&=1+\frac{\gamma-1}{2}M_{t}^{2}&& \\
    \frac{\rho_{c}}{\rho_{t}}&=\left(1+\frac{\gamma-1}{2}M_{t}^{2}\right)^{1/\left(\gamma-1\right)}&&
\end{flalign}
\endgroup

The velocity at the throat $V_{t}$ is related to the Mach number as $V_{t}=M_{t}\left(\gamma R T_{t}\right)^{1/2}$ whose sonic or subsonic behavior is determined by the choked flow condition. Finally, the temperature evolution $T_{C\text{\sout{\ensuremath{V}}},t+\Delta t}$ in any of the control volumes is predicted using the ideal gas law

\begingroup
\small
\begin{flalign}
    T_{C\text{\sout{\ensuremath{V}}},t+\Delta t}=\frac{p_{C\text{\sout{\ensuremath{V}}},t+\Delta t}}{\rho_{C\text{\sout{\ensuremath{V}}},t+\Delta t}R}&&
\end{flalign}
\endgroup

\begin{figure}
    \centering
    \includegraphics[width=0.5\textwidth]{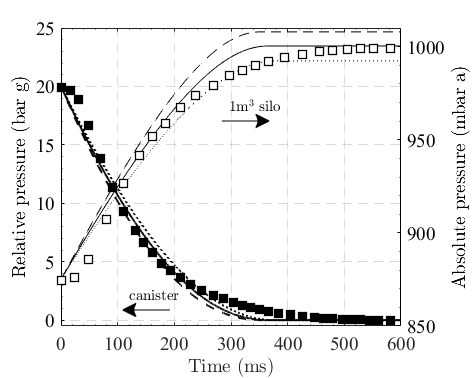}
    \caption{Pressure trends during the air blast: experimental data (markers) and poly-tropic model with $n=1.2$ (dotted lines), $n=1.3$ (solid lines), and $n=1.4$ (dashed lines).}
    \label{Fig:Pressure_trend_air_0Dmodel}
\end{figure}

Performance of the dispersion system was checked by
comparing the experimental measurements with the 0D poly-tropic model implemented in Matlab. After iterating over various poly-tropic exponents, the best agreement of the final pressure reading was found for $n=1.3$. The pressure trends in both the canister and 1 m\textsuperscript{3} silo are shown together with the model results in Fig. \ref{Fig:Pressure_trend_air_0Dmodel}. Although the ignition delay time was 600 ms, the experimental reading suggests that the pressure in the silo and canister stabilizes around 500 ms. In contrast, the time for stabilizing the pressures calculated by the model varies and is slightly ahead of the experimental data. Such offset can be attributed to various factors: (1) the spatial effects are not resolved (such as the length and curvature of the conduit), (2) friction effects are neglected, and (3) the time delay of the electropneumatic valve is ignored.\\

As illustrated in the figure, the pressure discharge in the experiment does not commence immediately. Instead, it appears to be delayed by approximately 20 ms before it gradually develops. To account for irrecoverable losses and improve the model prediction, a discharge coefficient $C_{d}$ could be introduced; however, this is difficult to estimate based on the number of parameters in the experiment. \\

Despite its simplicity, the poly-tropic model is a useful tool to get reasonable estimates of the behavior of the pressurized air injection. It helps to calculate the desired pressure upon commencing the explosion test and to establish an according ignition delay time. In addition, it can provide other useful information, such as the fact that in our current setting, the flow is highly turbulent for almost half of the air blast. During this period, the Reynolds number is in the order of $\mathcal{O}\left(Re\right)\sim10^{6}$ and the flow velocity becomes subsonic only after approximately 300 ms.\\

\subsection{Non-reactive flow}

In dust explosions the particle size is a crucial factor, as it influences not only the dynamic behavior of the dust cloud, but also the reactivity of the fuel. Thus, it is essential to analyze dust dispersion in order to fully comprehend the development of the explosion. Dust dispersion by pressurized air is the most common method for dust explosion testing. In this method, the dispersion nozzle plays a leading role in the whole dispersion system, as it distributes the dust inside the test vessel and regulates the flow pattern and turbulence intensity. In this work, the axisymmetric rebound nozzle produces a recirculating flow pattern characterized by four large vortices that emerge from the nozzle outlet and extend up to the top of the silo. These vortices are distributed symmetrically and rotate in clockwise direction (right vortex) and counter-clockwise direction (left vortex), as depicted by the streamlines in Fig. \ref{Fig:Velocity_Streamlines}. The top row of the figure corresponds to the dust-free flow injection, while the bottom row illustrates the behavior with a nominal dust concentration $C_{0}=500\,\text{g}\,\text{m}\textsuperscript{-3}$.\\

\begin{figure*}
\begin{center}
    \begin{tabular}{c|c}
    \multirow{2}{*}[5.2cm]{    \subfigure[Concentration Axial Lines \label{Fig:Concentration_axial_lines}]{    \includegraphics[width=0.6 \textwidth]{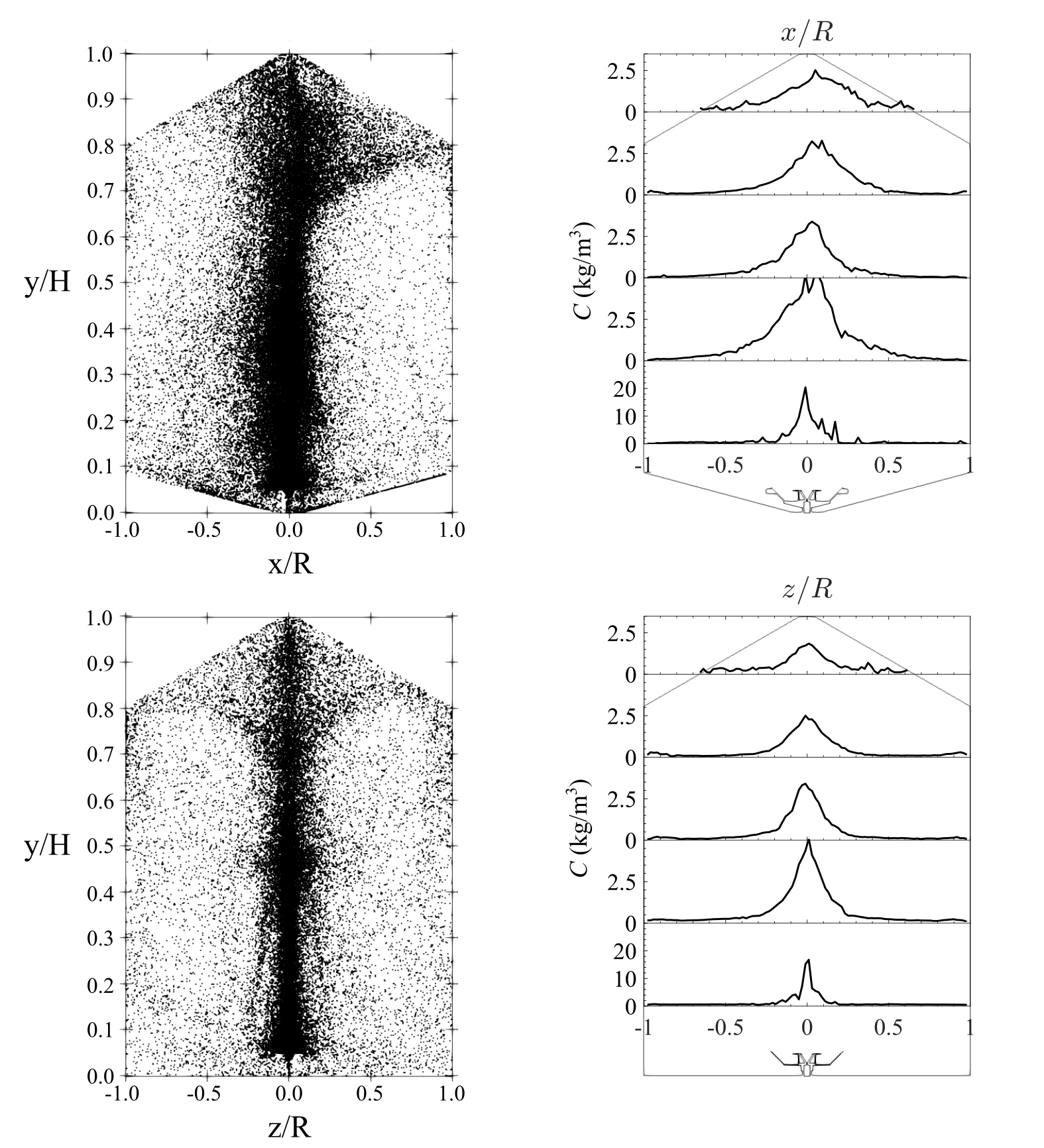}}} &     \subfigure[Stokes number \label{Fig:Stokes_number}]{\includegraphics[width=0.4\textwidth]{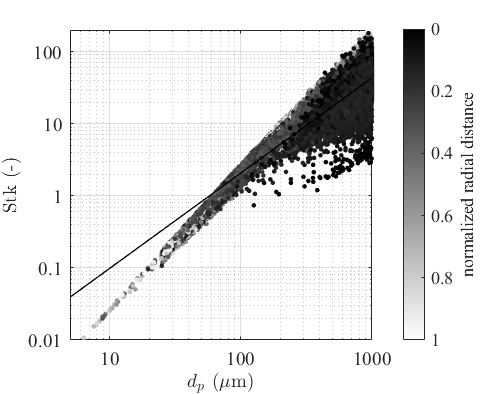}} \\
                  &     \subfigure[Reynolds number \label{Fig:Reynolds_number}]{\includegraphics[width=0.4\textwidth]{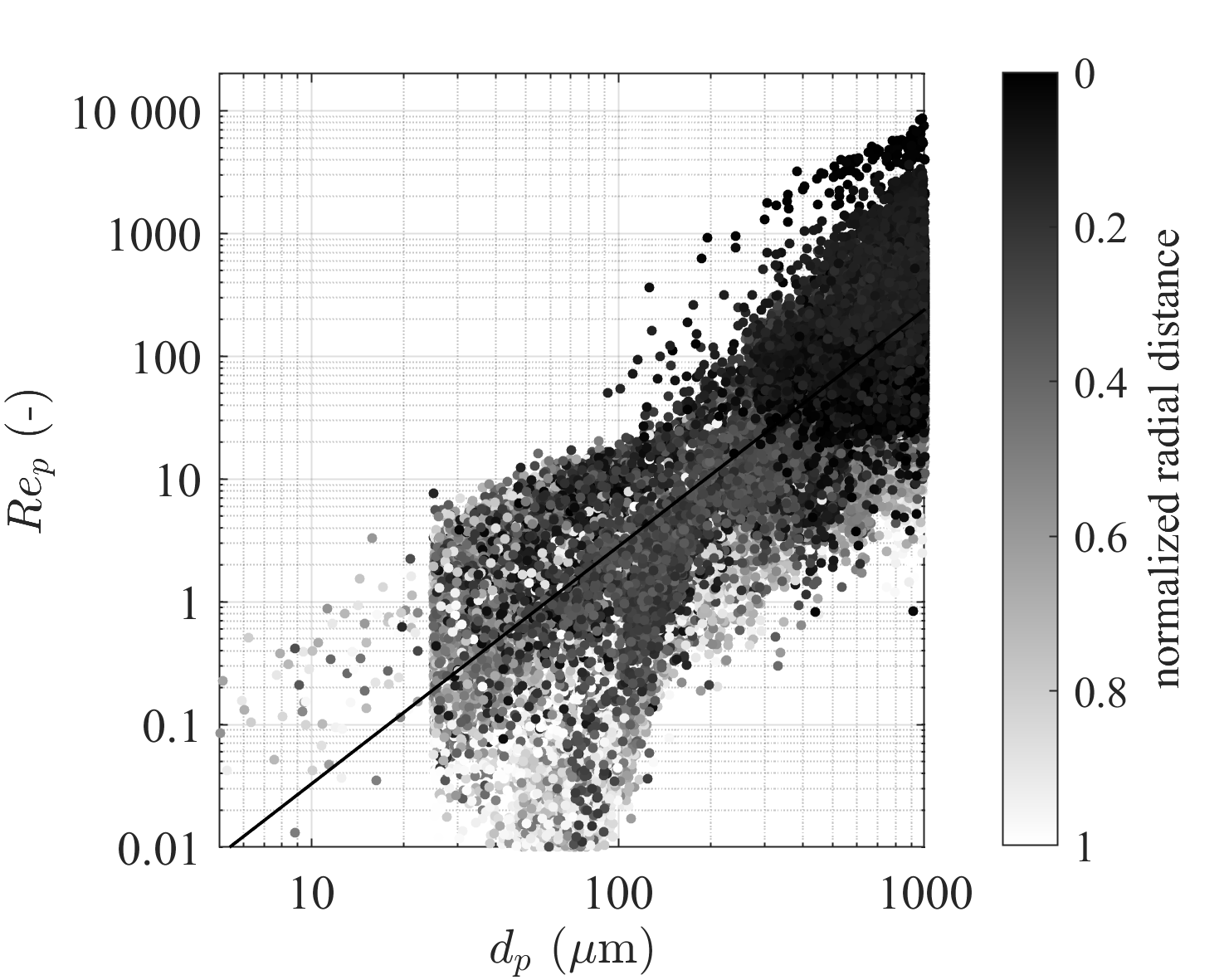}}    
    \end{tabular}
    \end{center}

\caption{Distribution, dust concentration and dimensionless numbers of the dust cloud prior ignition.}
\label{Fig:All_merged}
\end{figure*}

In all the experiments, the ignition delay time is 600 ms, and throughout the following discussion, all graphs and figures represent this time interval from -600 ms to 0 ms. In the early stage (as shown in Fig. \ref{Fig:Velocity_Streamlines_100ms}), the cores of the vortices are located at the top, specifically at a height of $0.6 < y/H < 0.8$, which is very close to the corners. When dust is injected, the vortices are still able to form, but their location is altered. The vortices shift downward to a height of approximately $y/H = 0.5$ for the same time interval. This occurs because the first injected particles have high velocities and collide strongly with the silo roof. As a result, they descend and hinder the vortices from extending all the way to the top.\\

As the flow progresses and particles are fully injected, the vortices rise vertically and the cores return to their original positions at the corners (refer to Fig. \ref{Fig:Velocity_Streamlines_600ms}). Moreover, by the end of the dispersion process, small vortices form at tip of the roof. Unfortunately, these recirculating flow structures do not facilitate dust cloud mixing, as they behave as stagnant zones, where only air is trapped. This flow pattern bears striking resemblance to that of the 20L Siwek sphere, where Benedetto et al. \cite{DIBENEDETTO2013cfd} exposed that the vortices tended to deposit most of the particles near the vessel walls. However, in case of the 1 m\textsuperscript{3} silo, the particles accumulate in the central section of the silo, particularly in a central column.\\ 

Fig. \ref{Fig:Concentration_axial_lines} depicts the particle distribution in two half-plane slices, revealing that the cloud is highly concentrated in the region between $-0.5 \leq x/R \leq 0.5$. The concentration in this central column varies locally along the height and is notably highest in the lower zone, where it reaches a maximum of $20 \,\text{kg}\,\text{m}\textsuperscript{-3}$ before decreasing to $2.5 \,\text{kg}\,\text{m}\textsuperscript{-3}$  in the uppermost part of the column.\\

When injecting a dust sample with large particle diameters, the particles may behave ballistically and interact very little with the air flow. A convenient way to determine if the particles trajectories adjust to the air streamlines is by calculating the Stokes number. The Stokes number is a parameter that relates the particle response time to a characteristic time scale of the fluid, which can be used to determine whether the particles are in equilibrium with the air or not. The Stokes number is:

\begingroup
\small
\begin{flalign}
\text{Stk}&= \frac{\tau_{p}}{\tau_{f}}&& \\
\tau_{p}&= \frac{\rho_{p}d_{p}^2}{18\mu_{f}}\frac{24}{C_{D}Re_{p}}&&
\end{flalign}
\endgroup

\noindent where $C_{D}$, $Re_{p}$ and $\tau_{f}$ are the particle drag coefficient, the particle Reynolds number, and the characteristic fluid time scale $\tau_{f}=l_{e}/|\tilde{u}_{i}|$, respectively. In these calculations, $l_{e}$ is the integral length scale $C_{\mu}k^{3/2}/\varepsilon$, the drag coefficient obeys the correlation shown in  Eq. (\ref{Eqn:Drag_coefficient}) and the particle Reynolds number is

\begingroup
\small
\begin{flalign}
    Re_{p}=\frac{|\tilde{u}_{i}-u_{p_{i}}|d_{p}}{\nu_{f}}&&
\end{flalign}
\endgroup

\begin{figure*}
      \begin{tabular}{c|c}
    \subfigure[Velocity fluctuations for the dust-free flow and particle-laden flow (left side) and instantaneous dust concentration in the 1m\textsuperscript{3} silo (right-side), where $C_{0}=500$ g/m\textsuperscript{3} is the nominal dust concentration \label{Fig:velocity_fluctuations}]{\includegraphics[width=0.50\textwidth]{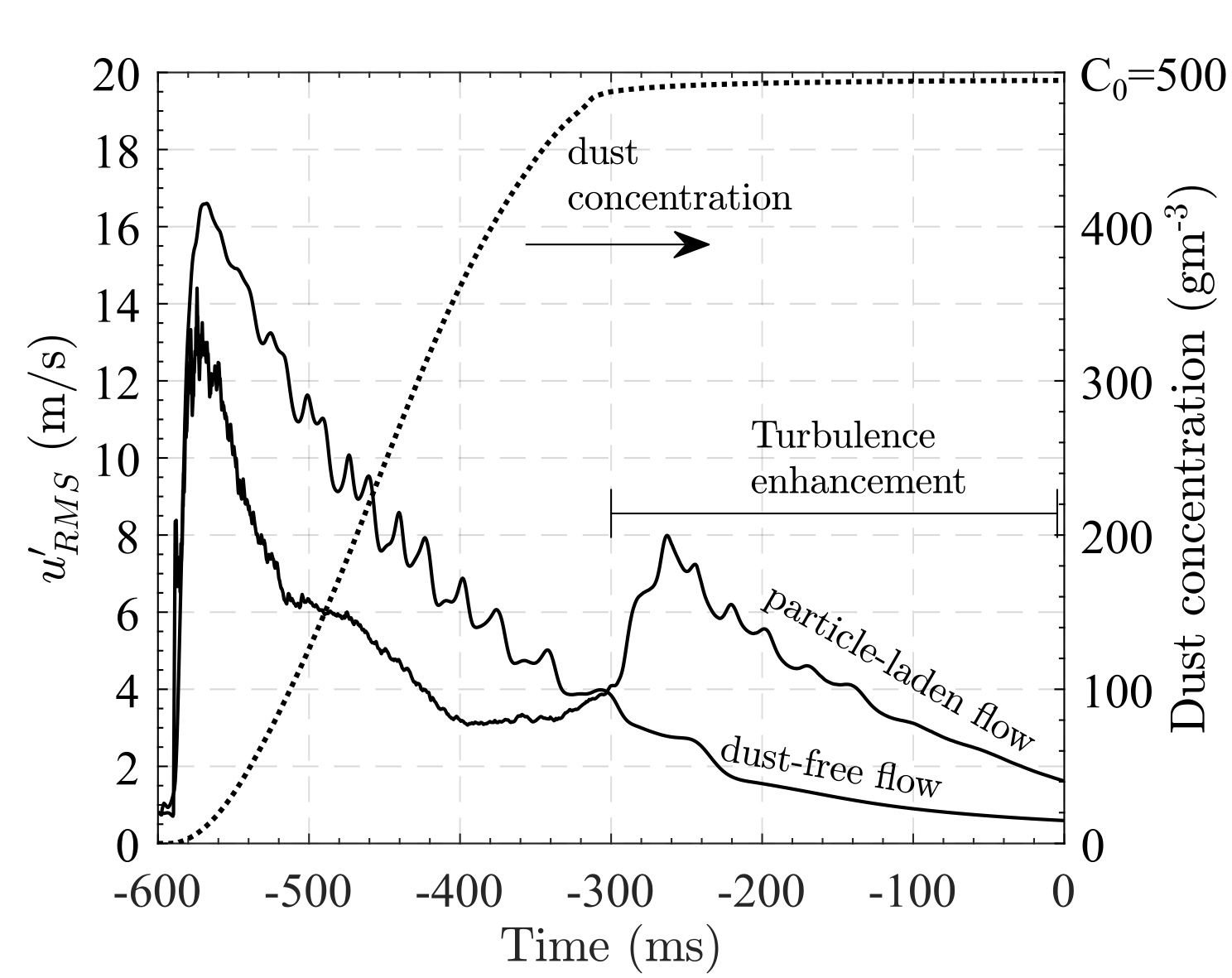}} & \subfigure[Percentage change in the turbulence intensity as a function of the ratio of particle size to Kolmogorov length scale $d_{p}/\eta$. Data represented as a Gore-Crowe diagram \cite{gore1989effect} \label{Fig:TurbModulation}]{\includegraphics[width=0.50\textwidth]{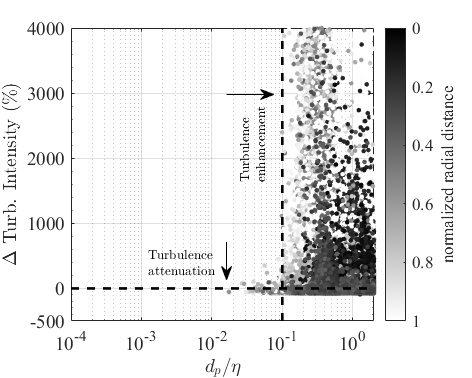}}
      \end{tabular}
      \caption{Some characteristics of the particle-laden flow compared to the dust-free flow}
      \label{Fig:Some_characteristics}
\end{figure*}

For particles with a Stokes number less than unity, the fluid can alter their trajectories and cause them to closely follow the motion of vortices. However, when the Stokes number of particles is greater than unity, their strong inertia impedes the fluid streamlines from altering their trajectories. Fig. \ref{Fig:Concentration_axial_lines} clearly shows that the cloud has a small radial aperture and very few particles are contained within the vortices. \\

Fig. \ref{Fig:Stokes_number} shows that the Stokes number logarithmically varies as a function of particle diameter, where a threshold value of $\text{Stk} = 1.0$ indicates a critical particle diameter of apparently 100 microns. Particles larger than this size display a response time up to two orders of magnitude greater than the fluid time scale, meaning that it takes significantly longer for them to adapt to vortex motion. Based on the particle size distribution depicted in Fig. \ref{Fig:PSD_comparison}, over 90\% of biomass dust particles are affected by this condition, as their adaptation time is over 100 times longer than the fluid time scale. Only particles with diameters less than 100 microns are able to interact with recirculating flow. Furthermore, the figure illustrates that the larger the particle diameter, the shorter its distance from the central dust column. Conversely, as particle diameter decreases, the particles can be dispersed over a wider range of radial distances.\\

In addition, Fig. \ref{Fig:Reynolds_number} shows that the particle Reynolds number also displays a logarithmic dependence on particle diameter, with particle inertia being up to four orders of magnitude greater than fluid inertia. This confirms that large particles exhibit a ballistic behavior and thus, the dust cloud is unable to mix homogeneously with the air.\\

Another important aspect of dust explosions is the initial turbulence at which the dust cloud ignites. Turbulence is a function of many aspects, such as the dispersion nozzle, the flow pattern, the dust concentration, the particle size or the ignition delay time. In explosivity tests in standardized vessels, it is well known that an increase in turbulence levels increases the rate of pressure rise. Fig. \ref{Fig:velocity_fluctuations} shows a comparison of the velocity fluctuations between the air-only case and the dust case.\\ 

For the air-only case, a maximum value of $u^{\prime}_{\text{RMS}}=16\,\text{m/s}$ is reached just about 40 ms after the start of the injection, whereafter the turbulence is strictly decreasing until the end of the injection delay time. This happens because the intensity of the pressure gradient decreases steeply and continuously until the pressures of both, the canister and the 1 m\textsuperscript{3} silo stabilize. In addition, the mechanisms of turbulence production such as wall friction and shear layers are not strong enough to counteract the decay of the pressure gradient. On the contrary, for the case with dust injection, there are two periods of turbulence decay. The first of these periods corresponds to the times between -600 and -400 ms, where the velocity fluctuations are smaller than in the case with air injection only. This is because the discharge of particle-laden air obstructs and slows down the flow entering the silo, debilitating the baroclinic force and decreasing the incoming velocity. The second period corresponds to the times between -260 and 0 ms, where the velocity fluctuations reached a maximum value of $u^{\prime}_{\text{RMS}}=8\,\text{m/s}$ and is higher than the case of injection with only air.\\

The intermediate time between -400 and -260 ms is a period of stabilization and turbulence production. Here, the decay of the first period is counteracted mainly for two reasons. The first is that by this time 70\% of the nominal dust concentration has already entered the 1 m\textsuperscript{3} silo, which frees the flow from local blockages caused by the particles. The second is that the particles already injected create local distortions in the flow, generating self-induced wakes and vortices around the particles. In this period, the velocity fluctuations double, rising from $u^{\prime}_{\text{RMS}}=4\,\text{m/s}$ to $8\,\text{m/s}$. \\

Interestingly, the turbulence of the particulate flow is consistently higher than that of the air-only flow from -300 to 0 ms, as shown in Fig. \ref{Fig:velocity_fluctuations}. This period is referred to as "turbulence enhancement". This finding is significant because the intensity of turbulence before dust cloud ignition can considerably impact the rates of heat transfer and chemical reactions during the explosion, a phenomenon known as turbulence modulation. Turbulence modulation is the inertial effect of particles on the flow turbulence relative to non-particle flow. Crowe \cite{crowe2011multiphase} identified some factors that influence turbulence modulation: surface effects, inertial effects, and response effects.\\

\begin{figure}[h]
    \centering
    \includegraphics[width=0.5\textwidth]{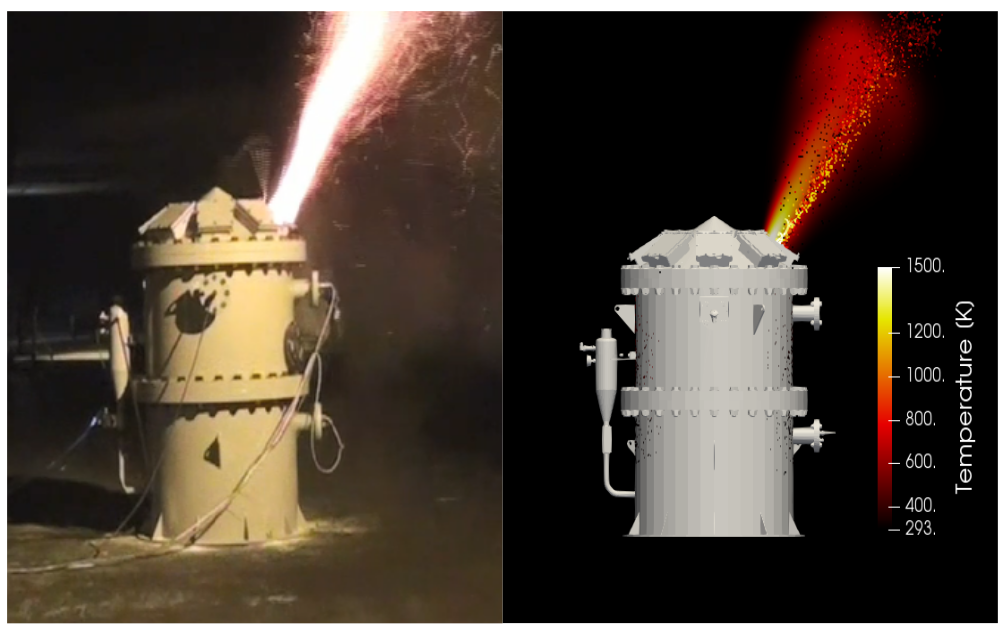}
    \caption{Qualitative comparison of the flame propagation between the explosion experiment (left) and the CFD simulation (right).}
    \label{Fig:Explosion_Experiment}
\end{figure}

On the one hand, the inertial and response effects are determined by the dimensionless numbers mentioned earlier. Because the particle Reynolds number and Stokes number cover several orders of magnitude above unity, the dispersion of this biomass sample is likely to significantly affect the flow turbulence compared to the flow without particles.\\

On the other hand, surface effects are determined by the particle diameter normalized by a length scale representative of the flow. As the dispersion time comes to a close, turbulence decreases for both dust-free flow and particle-laden flow. The Kolmogorov scale, which represents the smallest eddies where viscosity is the dominant force and turbulent kinetic energy is dissipated in heat, serves as a characteristic length scale for turbulent eddies. Fig. \ref{Fig:TurbModulation} shows the change in turbulence intensity versus particle diameter normalized by the Kolmogorov length scale. The percentage change in turbulence intensity is defined as

\begingroup
\small
\begin{flalign}
    \Delta \text{Turb. Intensity \%}=\frac{\text{Turb. Int.}\textsubscript{TP}-\text{Turb. Int.}\textsubscript{SP}}{\text{Turb. Int.}\textsubscript{SP}}\times 100&&
\end{flalign}
\endgroup

\noindent where the turbulent intensity is calculated based on the hypothesis of isotropic turbulence, $\text{Turb. Int}=\sqrt{2k/3}/|\tilde{u}_{i}|$ and the subscripts TP and SP refer to the two-phase and single-phase flows, respectively. All values were calculated locally at the particle positions inside the 1 m\textsuperscript{3} silo.\\

The Gore-Crowe classification identifies the critical value $d_{p}/\eta = 0.1$, which marks the threshold at which higher values of $d_{p}/\eta$ will cause an increase in the turbulence intensity of the entrained gas, and lower values will cause a decrease. The figure shows that the particles generally follow this criterion, with most of the data points lying to the right of the threshold value and above the horizontal line representing zero. Notably, particles located closest to the central dust column exhibit the highest $d_{p}/\eta$ ratios.\\

A possible explanation for this phenomenon is that the particles generate turbulence in their wake at the length scale of the smallest eddies, resulting in an increase in the turbulence intensity of the air. In this case, the energy is transferred from the particles to the turbulent kinetic energy of the trailing gas. Therefore, according to the map, when injecting a dust loading of $500\,\text{g}\,\text{m}\textsuperscript{-3}$, the local increase in turbulence intensity can be as high as 4000\%.\\

\subsection{Reactive flow}

\begin{figure}[h]
    \centering
    \includegraphics[width=0.5\textwidth]{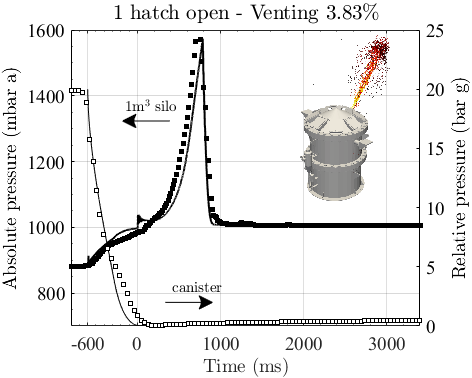}
    \caption{Vented dust explosion (venting 3.83\%): comparison of the pressure-time histories between the experimental test results (markers) and the CFD simulation (solid lines).}
    \label{Fig:Vented_explosion_1hatch}
\end{figure}

The venting experiments were conducted during the last quarter of year 2022 in the experimental test tunnel of Applus+ TST. A delimiting explosive area was established and the 1 m\textsuperscript{3} silo was safely installed and anchored to the ground. The tests were carried out under the supervision and support from IDONIAL and Applus+ TST personnel and were operated from a remote control unit. All tests were performed for a concentration of $500\,\text{g}\,\text{m}\textsuperscript{-3}$ with an ignition delay time of 600 ms and an activation energy of 10 kJ. Fig. \ref{Fig:Explosion_Experiment} shows a qualitative comparison of the flame propagation between the experiment and the simulation during a vented explosion with 1 hatch open (venting percentage of 3.83\%).\\

Upon the rupture of the polymer bolt, the hinged hatch opens violently and lets a jet flame to escape 
perpendicularly to the silo roof. The gaseous flame comes out accompanied with burning biomass particles and an elevated inertia. Remarkably, the CFD simulation predicts a flame shape that resembles very well to the experimental observation, with a flame temperature that is close to 1500 K and that extinguishes after a few seconds.\\

Fig. \ref{Fig:Vented_explosion_1hatch} displays the pressure profile recorded in the 1 m \textsuperscript{3} silo and canister during the test. The graph illustrates the pressure increase in the silo on the left side and the pressure discharge in the canister on the right side. As noted, the pressure profile predicted by the CFD model agrees very well with the experimental test, both showing a maximum pressure value of 1570 mbar a. This value corresponds to the rupture pressure of the polymer bolts, which triggers the opening of the hinged hatch, allowing the pressure wave and flame to escape from the silo to the surroundings. Shortly after, the pressure inside the silo drops abruptly to atmospheric pressure.\\

A key metric of the test is the time taken to vent the explosion. The CFD model predicts that the static pressure of the bolts is reached 785 ms after activating the igniters, while the experimental test records a time of 757 ms. The relative error is 3.5\% and endorses that the model is in excellent agreement with the experiment, not only capturing the maximum overpressure or the flame propagation, but the transient behavior in general.\\

\begin{figure}
    \centering
    \includegraphics[width=0.5\textwidth]{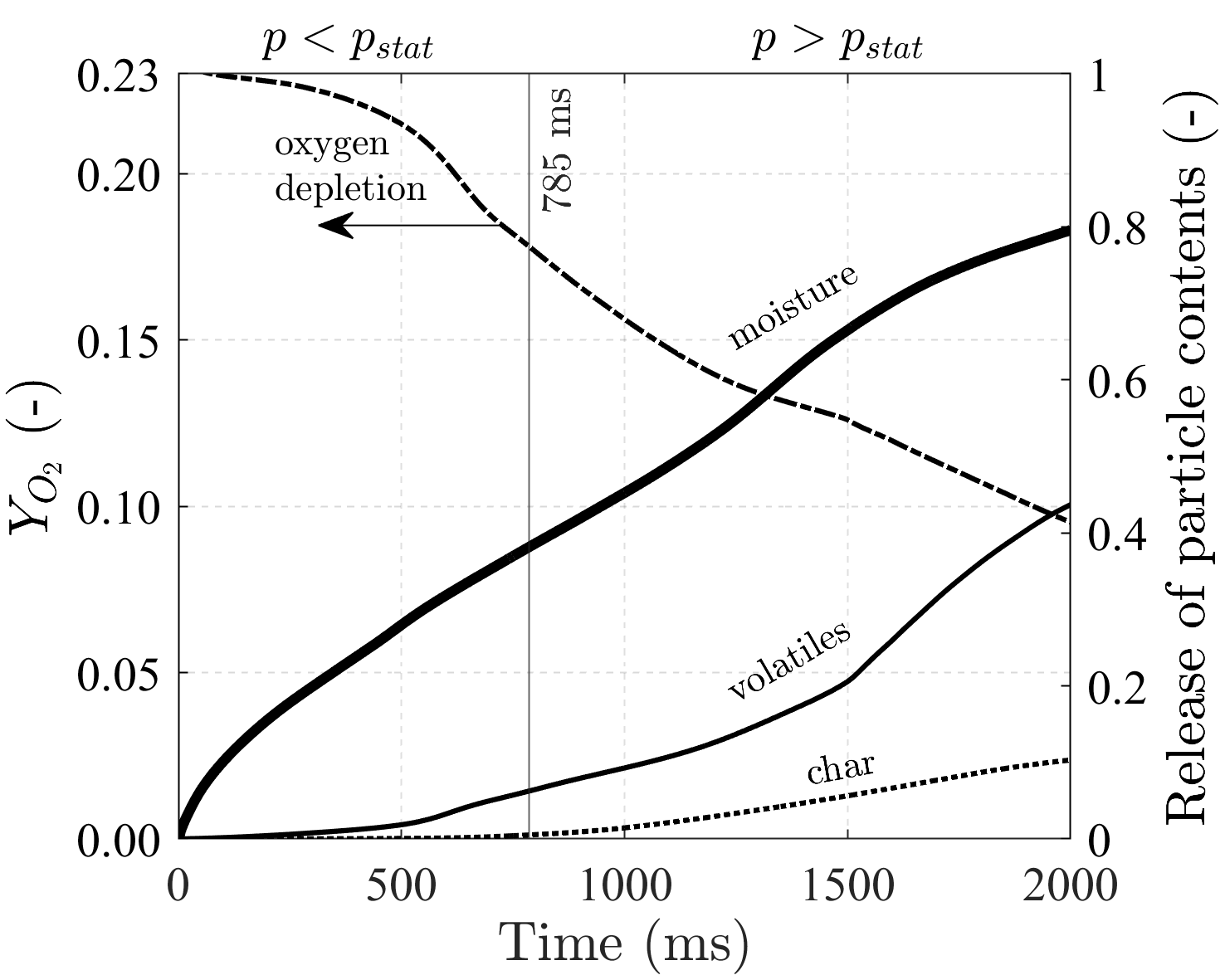}
    \caption{Temporal evolution of the release of particle contents and oxygen depletion during the dust explosion.}
    \label{Fig:Particle_mass_yields}
\end{figure}

To quantify the fuel burned during the explosion, we analyzed the consumption of each reactive component in the dust cloud, as shown in Fig. \ref{Fig:Particle_mass_yields}. The graph is divided into two regions: (1) before the opening of the hatch ($p < p_{stat}$) and (2) after the opening of the hatch ($p > p_{stat}$). The data in the first region of the graph shows that prior the opening of the hatch, the O\textsubscript{2} mass fraction decreased from 0.23 to 0.17. This suggests that the dust cloud burned in small amounts, with only approximately 10\% of the volatile gases being released from the particles. Despite this, the combustion of such small amount of gases was enough to create an overpressure of 570 mbar g inside the silo.\\ 

\begin{figure*}
    \centering
    \includegraphics[width=0.8\textwidth]{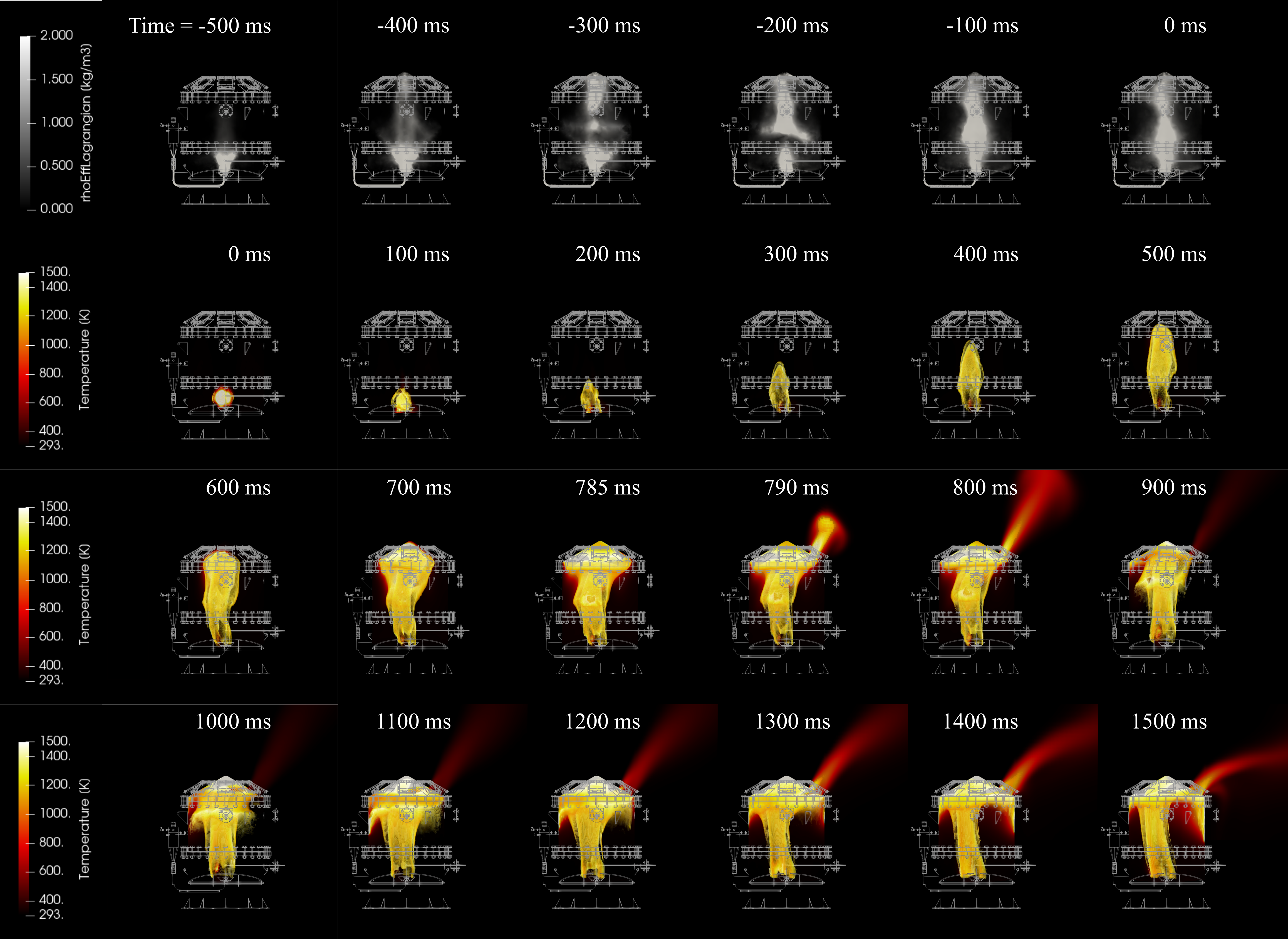}
    \caption{Evolution of the dust explosion with 1 hatch open (venting 3.83 \%). Contours of particle concentration (dust dispersion) and contours of temperature (closed and vented explosion).}
    \label{Fig:Contours_1_Hatch_open}
\end{figure*}

In the second region of the graph, it is evident that the other reactive components of the cloud burned slowly, with a slight increase in the consumption rates starting at 1500 ms. However, even after 2 seconds, only 40\% of the volatile gases had been released and significant amounts of moisture and char remained in the particles. Based on the proximate analysis of the biomass dust, only 50 g of the available 500 g of mass had been consumed by the time the hatch opened, and only 194 g had been consumed after 2 s of ignition.\\

Fig. \ref{Fig:Contours_1_Hatch_open} illustrates the evolution of the dust explosion during all the stages of the experiment. The first 6 contours represent the dust injection phase, which is non-reactive flow. These contours are colored by dust concentration and are spaced every 100 ms until the time 0 is reached. From that point, the temperature contours of the reactive flow are displayed up to 1500 ms.\\

The flame development begins with the activation of the pyrotechnic igniters, which, as mentioned before, are modeled as a 10kJ sphere of radius 13 cm placed above the axisymmetric rebound nozzle. These igniters generate the initial flame that induces the dust particles to produce a self-propagating flame kernel. The resulting flare propagates vertically, creating a mushroom-shaped flame. This is primarily due to 2 factors:

\begin{figure}[h]
    \centering
    \includegraphics[width=0.5\textwidth]{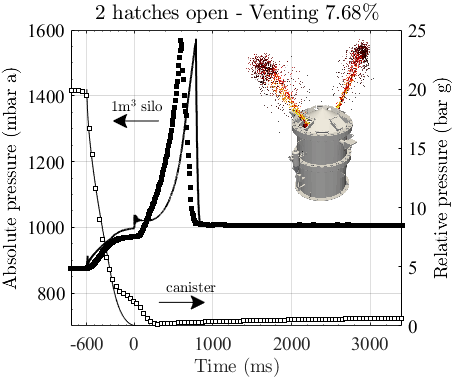}
    \caption{Vented dust explosion (venting 7.68\%): comparison of the pressure-time histories between the experimental test results (markers) and the CFD simulation (solid lines).}
    \label{Fig:Vented_explosion_2hatch}
\end{figure}

\begin{itemize}
    \item First, hot gases are drawn upward by the flow pattern and velocity field achieved at the end of the dispersion process. This is evident from Fig. \ref{Fig:Velocity_Streamlines}.
    \item Second, the flame spreads in the direction where the fuel is present, which is linked to the distribution of the dust cloud during the injection phase. As discussed in the previous section, the large particle size of the dust resulted in a thin central cloud distribution.
\end{itemize}

Once the flame reaches the uppermost part of the silo, it hits the roof and expands radially, seeking areas where the local equivalence ratio allows for stoichiometric combustion. The mushroom appearance is due to the fact that the local dust concentration is higher near the silo roof than within the vortex cores, where the dust/air mixture is extremely lean, as shown in Fig. \ref{Fig:Concentration_axial_lines}.\\

Starting at 785 ms, the temperature contours correspond to the vented explosion. The jet flame can be observed escaping perpendicular to the roof and reaching its maximum length and temperature within the next 100 ms. This phenomenon can be attributed to the pressure gradient that propels the flame from the silo to the surrounding atmosphere. Once the pressure in the silo stabilizes with atmospheric pressure, the jet flame partially extinguishes. After 1000 ms, the flame is reignited due to the unburned particles that left the silo and encountered fresh oxygen in the atmosphere. However, the flame weakens and bends down since the velocity magnitude of the jet has decayed significantly.\\

\begin{figure*}
    \centering
    \includegraphics[width=0.8\textwidth]{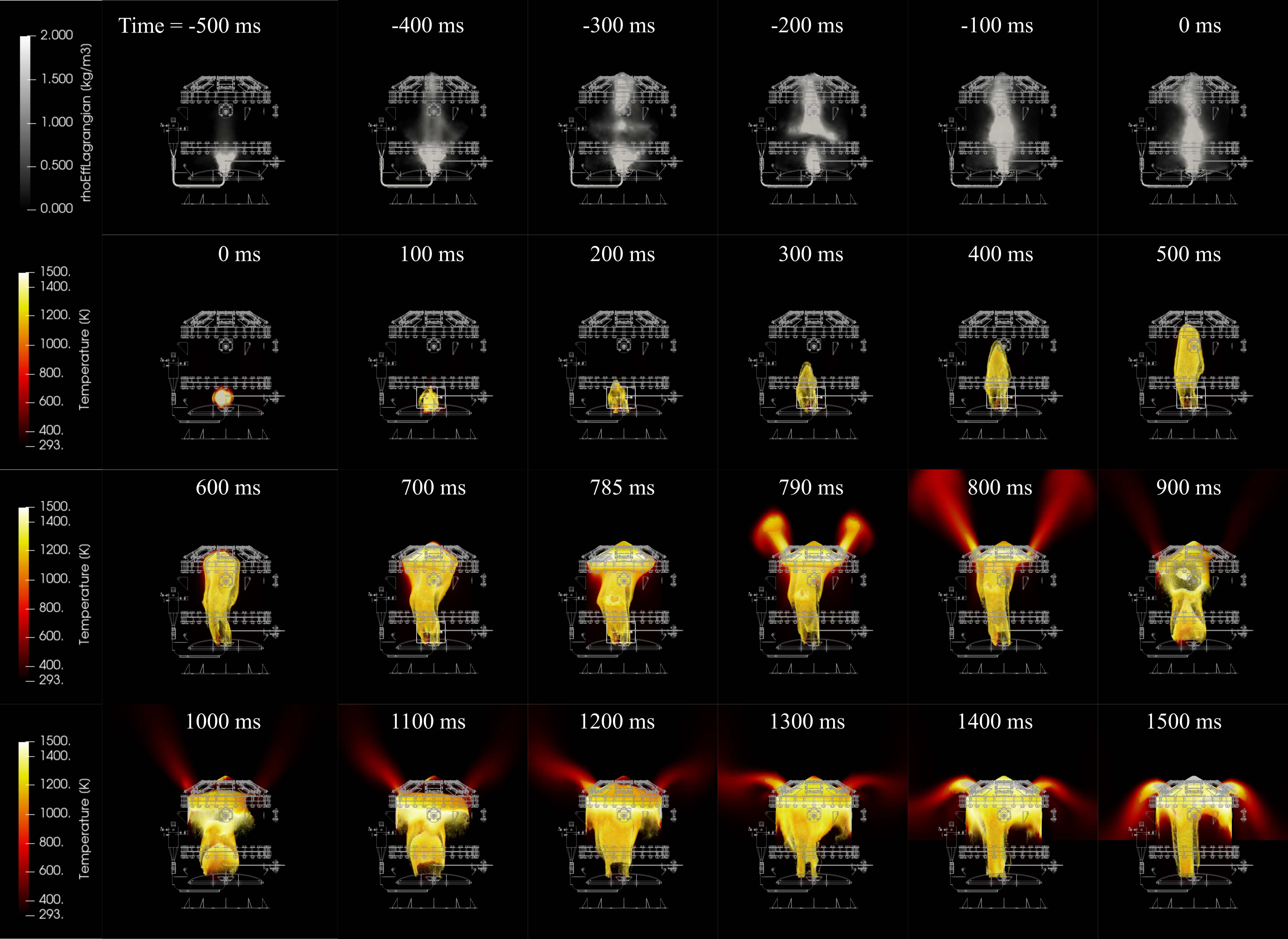}
    \caption{Evolution of the dust explosion with 2 hatches open (venting 7.68 \%). Contours of particle concentration (dust dispersion) and contours of temperature (closed and vented explosion).}
    \label{Fig:Contours_2_Hatches_open}
\end{figure*}

An additional experimental test was conducted with two open hatches and identical operating conditions as the previous test. Pressure curves are shown in Fig. \ref{Fig:Vented_explosion_2hatch}. Again, once the rupture pressure of the bolts is reached, the pressure in the silo decreases rapidly until it reaches atmospheric pressure. However, this time the CFD model is slightly delayed compared to the experimental results, with the static pressure of the bolts being reached at 785 ms in the simulation compared to 600 ms in the test. This can be attributed to an unexpected incidence with the experimental test. According to the pressure discharge in the canister, the experimental reading registered a bump right below the 5 bar g and during almost the second half of the ignition delay time. This delay may have occurred due to a dust blockage in the dispersion duct, which slowed pressure balancing in both vessels. A weakened pressure gradient may cause the initial pressure in the silo to decrease at the moment of dust cloud ignition.\\

Igniting the air/dust mixture at pressures below 1 bar may affect mass transfer rates, particularly the evaporation rates due to the pressure-dependent nature of water's phase change from liquid to vapor state. At pressures below 1 bar, the evaporation point of moisture is reduced, so during the first 100 ms of the experimental test, the dust cloud may have burned faster than if it were ignited at atmospheric pressure. This faster evaporation of moisture may result in the earlier release of volatile gases, advancing the pressure rise curve. Nevertheless, the CFD model predicts transient effects quite accurately.\\

Fig. \ref{Fig:Contours_2_Hatches_open} illustrates the evolution of this experiment. As only the number of open hatches was changed for this test, all contours up to 785 ms are identical to those of the previous explosion. Two flame jets escape perpendicularly to the silo roof, developing their maximum length and temperature within a few milliseconds after opening. The flames partially extinguish after 900 ms and appear again after 1100 ms. The first jet flame is a consequence of the pressure wave that drags the hot gases and particles towards the far field, while the second flame is due to the fact that both the particles and the fresh gases escaping from the silo encounter abundant oxygen in the surroundings. The forming flames attach to the periphery of the hatches and bend downward because of a debilitated velocity field. Also, it is interesting to note that the flame inside the silo maintains its mushroom shape and does not propagate into the interior of the vortices. This suggests that the particles never mixed with the recirculating flow pattern and remained distributed in the central column previously studied.\\

\begin{figure*}
      \begin{tabular}{c|c}
    \subfigure[Volatile yield as function of particle diameter at the time of the hatch opening. Instantaneous data (points) and moving average (solid line) \label{Fig:Volatile_Yield}]{\includegraphics[width=0.50\textwidth]{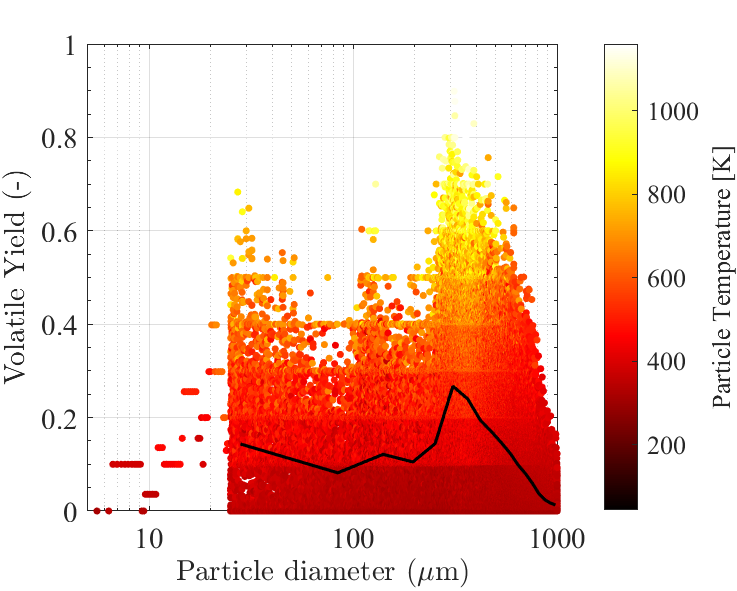}} & \subfigure[Char burnout as function of particle diameter at the time of the hatch opening. Instantaneous data (points) and moving average (solid line) \label{Fig:Burnout}]{\includegraphics[width=0.50\textwidth]{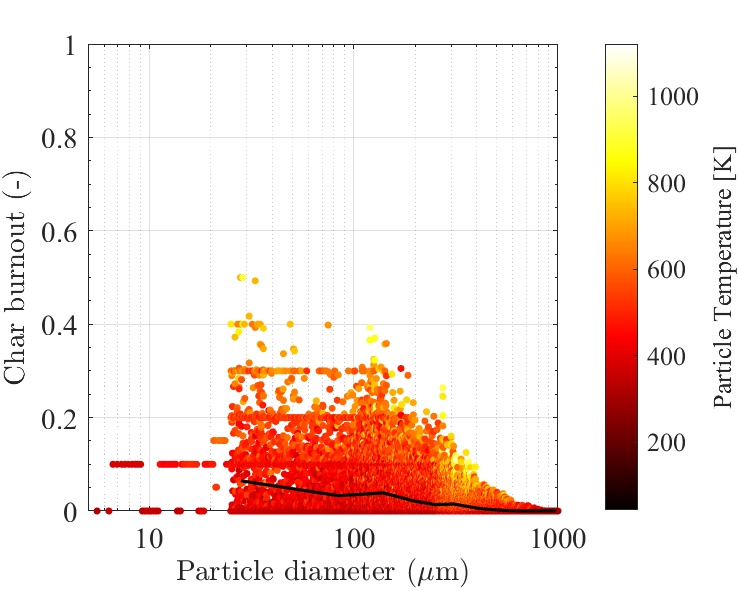}}
      \end{tabular}
      \caption{Reactivity of biomass particles as function of particle diameter}
      \label{Fig:Volatiles_Burnout}
\end{figure*}

Finally, we identified the particle size with higher reactivity at the time of hatch opening, Fig. \ref{Fig:Volatiles_Burnout} classifies the consumption of each fuel component based on particle diameter. The amount of gases released from the dust cloud is higher than the burned mass of the fixed char, as expected due to the higher volatile matter content in biomass combustion. Furthermore, particles with a diameter close to $300\,\mu\text{m}$ not only reached the highest temperatures, but also exhibited the highest reactivity with respect to devolatilization. While this may seem counter-intuitive, in our previous works \cite{ISLAS2022117033,islas2022computational}, we have consistently emphasized that a dust explosion, specifically flame propagation, cannot be attributed solely to isolated factors such as particle size, dust distribution, velocity field, or residence time. Instead, it is the combined effect of all these factors that drives the phenomena. Within our 1m\textsuperscript{3} silo, we have demonstrated the non-uniform dispersion of dust, with particles concentrated in a central column, particularly larger particles (>100 $\mu$m) that represent the majority of the particle size distribution (as shown in Fig. \ref{Fig:PSD_comparison}). Additionally, particles smaller than 100 $\mu$m exhibit some radial scattering, as depicted in Fig. \ref{Fig:All_merged}. Consequently, it is the larger particles that ignite the air/dust mixture in our specific case, contradicting conventional expectations. This counter-intuitive behavior arises because the larger particles are aligned with the ignition source and effectively convect the flame vertically upward, as evidenced in Fig. \ref{Fig:Velocity_Streamlines} in conjunction with either Fig. \ref{Fig:Contours_1_Hatch_open} or Fig. \ref{Fig:Contours_2_Hatches_open}.\\

A notable distinction arises when comparing the flow characteristics achieved in apparatuses like the Godbert Greenwald (G-G) furnace to that of our 1m\textsuperscript{3} silo. The (G-G) furnace, often used for pyrolysis studies at high-heating rates \cite{pietraccini2023study}, exhibits a flow with relatively lower turbulence levels and enhanced uniformity due to its simplified geometry and absence of dispersion nozzles. 
The small pressure gradient used for dust dispersion in the (G-G) furnace leads to a flow environment that is potentially more uniform. This results in a lower level of turbulence and enhanced consistency in terms of velocity field, temperature, and dust distribution. As a consequence, all particles, irrespective of their size, have the opportunity to react under similar conditions, making it easier to establish correlations between devolatilization times and particle sizes. In contrast, our 1m\textsuperscript{3} silo experiences significantly higher turbulence and attains sonic velocities. The complex dust dispersion pattern observed within our silo further contributes to the non-homogeneous nature of the combustion process. Therefore, \textit{“turbulent dust flames depend on the nature of the turbulent flow field and thus on the experimental apparatus and are not basic to the dust itself”} as suggested by Smoot \cite{kramer1988oxidation}.


\section{Conclusions}
\label{Section:Conclusions}

In this study we conducted experimental tests and CFD simulations of biomass dust explosions in a newly developed 1 m\textsuperscript{3} silo apparatus designed for analyzing explosions in situations with variable venting. We examined all stages of a biomass dust explosion, including dust dispersion, ignition, closed and vented explosion. Our CFD results indicate that the flow characteristics after dust dispersion plays a crucial role in flame propagation and the explosion itself, and depends largely on particle size and the dispersion system. The turbulence prior to ignition and distribution of the dust particles also significantly affect the reactive characteristics of the cloud. During the explosion, our CFD model accurately predicted the time evolution of the pressure, particularly with regard to maximum overpressure and pressure relief. We observed similar pressure drops for the two venting scenarios studied.\\

The promising results obtained from our CFD simulations encourage the use of our CFD model to simulate larger scale geometries for further investigation of dust explosions. Future work will involve simulating additional test cases to gain a deeper understanding of the explosion behavior of biomass dust, especially in venting situations that fall out of the scope of the NFPA 68 or EN 14491 standards, and to help design effective safety measures to prevent such incidents.


\section*{Declaration of Competing Interest}
The authors declare that they have no known competing financial interests or personal relationships that could have appeared to influence the work reported in this paper. \\

\section*{CRediT Authorship Contribution Statement}

\textbf{A. Islas:} Conceptualization, Formal analysis, Data curation, Methodology, Software, Validation, Investigation,  Resources, Writing - original draft, Writing - review \& editing, Visualization. \textbf{A. Rodríguez Fernández:}  Methodology, Software, Validation, Investigation,  Resources, Writing - review \& editing. \textbf{E. Martínez-Pañeda:} Conceptualization, Writing - review \& editing, Funding acquisition. \textbf{C. Betegón:} Writing - review \& editing, Supervision, Project administration, Funding acquisition. \textbf{A. Pandal:} Conceptualization, Methodology, Software, Investigation, Resources, Writing - review \& editing, Supervision, Funding acquisition. \\

\section*{Acknowledgements}

Authors acknowledge that this work was partially funded by CDTI (Centro para el Desarrollo Tecnológico Industrial de España, IDI-20191151), Universidad de Oviedo and \linebreak PHB WESERHÜTTE, S.A., under the project "FUO-047-20: Desarrollo de silo metálico de grandes dimensiones ante los condicionantes de explosividad de la biomasa". Likewise, authors endorse the computer resources provided in the Altamira Supercomputer at the Institute of Physics of Cantabria (IFCA-CSIC), member of the Spanish Supercomputing Network, and the technical support provided by the Advance Computing group at University of Cantabria (UC) (RES-IM-2022-3-0002). A. Islas acknowledges support from the research grant \#BP20-124 under the 2020 Severo Ochoa Pre (Doctoral) Program of the Principality of Asturias.\\


\bibliographystyle{unsrt_abbrv_custom}

\bibliography{cas-refs}





\end{document}